\documentclass[twocolumn,
  amsmath,amssymb,amsfonts, a4paper,aps,
  citeautoscript,superscriptaddress,footinbib,prb]{revtex4-2}
\usepackage{txfonts}
\usepackage{color}
\usepackage{url}

\usepackage[colorlinks,bookmarks=false,citecolor=blue,linkcolor=red,urlcolor=blue]{hyperref}

\usepackage{graphicx}
\usepackage{dcolumn}
\usepackage{bm}
\usepackage[version=4]{mhchem}
\usepackage{subfigure}
\usepackage{verbatim}
\usepackage{booktabs}
\usepackage{braket}
\usepackage{hyperref}
\usepackage{amsfonts}
\usepackage{color}
\usepackage{url}

\usepackage[colorlinks,bookmarks=false,citecolor=blue,linkcolor=red,urlcolor=blue]{hyperref}

\begin{document}


\title{Ground State Phase Diagram of $\text{SU}(3)$  $t$-$J$ Chain}

\author{Junhao Zhang}
 \email{jhzhang16@fudan.edu.cn} %
\affiliation{Department of Physics and State Key Laboratory of Surface Physics, Fudan University, Shanghai 200433, P.R. China}
\author{Jie Hou}%
\affiliation{Department of Physics and State Key Laboratory of Surface Physics, Fudan University, Shanghai 200433, P.R. China}
\author{Jie Lou}%
 \email{loujie@fudan.edu.cn}
\affiliation{Department of Physics and State Key Laboratory of Surface Physics, Fudan University, Shanghai 200433, P.R. China}
\author{Yan Chen}
 \email{yanchen99@fudan.edu.cn}
\affiliation{Department of Physics and State Key Laboratory of Surface Physics, Fudan University, Shanghai 200433, P.R. China}
\affiliation{Shanghai Branch, Hefei National Laboratory, Shanghai 201315, P.R. China}

\date{\today}

\begin{abstract}
Distinct from the $\text{SU}(2)$ case, the fermionic systems with  $\text{SU}(N)$ symmetry are expected to exhibit novel physics, such as exotic singlet formation.
Using the density matrix renormalization group technique, we obtain the ground state phase diagram of the $\text{SU}(3)$ $t$-$J$ chain for density $n<1$.
The ground state phase diagram includes the Luttinger liquid, the extended Luther-Emery liquid characterized by a spin gap, and the phase separation state. 
We quantitatively assess the characteristics of the three phases by measuring spin gap, compressibility, various correlation functions, and structure factors.
We further study the extended Luther-Emery liquid phase and discover molecular superfluid quasi-long-range order.
The mechanism of the molecular superfluid is the combination of three $\text{SU}(3)$ fermions on sites that are not entirely connected.
Accordingly, we can speculate the behavior of the $\text{SU}(N)$ $t$-$J$ chain model with larger $N$ values, operating within the same filling regime.

\end{abstract}

\maketitle


\section{\label{Intro}Introduction}

The mechanism of high-$T_c$ superconductivity and strongly correlated electron systems has attracted great interest in both theoretical and experimental studies\cite{bednorz_1986,anderson_1987}.
The two-dimensional \emph{t-J} model is generally considered a minimal model that captures the essential interplay between charge and spin degrees of freedom in cuprates\cite{zhang_1988}.
It is the effective Hamiltonian of the Hubbard model in the limit of strong correlation.
Meanwhile, the one-dimensional (1D) \emph{t-J} model exhibits several exciting properties. 
The ground state phase diagram of the 1D \emph{t-J} model was first demonstrated by M. Ogata \emph {et~al.} \cite{ogata_phase_1991} theoretically and later verified by A. Moreno \emph {et~al.}\cite{moreno_ground-state_2011} later with density matrix renormalization group (DMRG) algorithm.
The phase diagram contains Luttinger liquid regions with the behavior of repulsive and attractive (i.e., superconducting) nature, spin gap, and phase separation for different values of $J/t$. 
The correlation functions and energy gaps determine their phase boundaries.

The spin freedom of electrons contains two spin components that expand the space carrying the fundamental representation of $\text{SU}(2)$.
A generalized symmetry $\text{SU}(N)$ has been proposed in different fermionic systems. 
One of the most famous examples is the role $\text{SU}(3)$ gauge symmetry playing in quantum chromodynamics\cite{gross_1973,politzer_1973}.
Recently, the theoretical exploration of the twisted bilayer graphene also leads to $\text{SU}(4)$ symmetry for the multiple of electron spin and twofold valley degeneracy\cite{yuan_2018}.


The development of quantum simulators provides a new approach to studying theoretical models through experimental measurements\cite{bloch_ultracold_2005,blatt_quantum_2012,bloch_quantum_2012,zhang_controlling_2020}. 
Ultracold Fermi gases of alkaline-earth isotopes, such as \ce{^{173}Yb} and \ce{^{87}Sr}, can carry multiple components with an intrinsic degree of freedom by their nuclear spins $I$. 
Interaction between atoms in optical lattices can be modulated by optical Feshbach resonance to have $\text{SU}(N)$ symmetry. 
For relatively small values of $N$, $\text{SU}(N)$ symmetry can be achieved by selecting nuclear spin states\cite{tusi_flavour-selective_2021}. 
Moreover, experimentalists have developed the ability to measure correlations in ultracold \ce{^{173}Yb} $\text{SU}(6)$ Hubbard model in the Mott insulator regime\cite{taie_su6_2012,taie_observation_2022}. 
These advances shed light on exploring many-body physics with $\text{SU}(N)$ symmetry. 
Systems with $\text{SU}(N)$ symmetry instead of $\text{SU}(2)$ spin symmetry may lead to the emergence of novel quantum phases and nontrivial physical behaviors. 

The zero-temperature phase diagram of the 1D $\text{SU}(N)$ Fermi-Hubbard model with $N>2$ has been investigated by various analytical and numerical techniques. 
The results are sketched for both incommensurate and commensurate fillings\cite{capponi_phases_2016}.
For incommensurate fillings, a metallic state shows up.
Its physical property strongly depends on the sign of the coupling constant $U$.
When $U>0$, all modes are gapless, and a metallic Luttinger liquid phase emerges\cite{gogolin_bosonization_1998,giamarchi_quantum_2004}.
On the other hand, a spin gap for $\text{SU}(N)$ degrees of freedom develops in an attractive case ($U<0$).
The charge density wave (CDW) or molecular superfluid (MS) instability dominates for high-density or low-density regimes.
For commensurate fillings, the ground state tends to be $p$-merization with gapped charge mode and gapless spin mode when the filling equals $1/p$ ($p$: integer).

The 1D $\text{SU}(N)$ \emph{t-J} model with a density below 1 (equivalently below $1/N$ filling) was studied by P. Scholttman\cite{schlottmann_groundstate_1993} as an exactly solvable model at the supersymmetric line $t=J$ with $\text{SU}(N|1)$ symmetry.
The global phase diagram of the 1D $\text{SU}(N)$ \emph{t-J} model away from the supersymmetric line is yet to be thoroughly investigated.
Whether the constraint on the local occupancy in strong coupling limit, which forbids the direct interaction between fermions of the number $N$ $(N>2)$, can break the emergence of the $\text{SU}(N)$ singlet attracts our interests. 
At the particle density $n=1$, $\text{SU}(N)$ \emph{t-J} chain model turns into 1D $\text{SU}(N)$ Heisenberg model with the fundamental representation (represented by Young diagram $\Box$) on each site.
Theoretical studies using conformal field theory (CFT)\cite{itoi_phase_2000,affleck_exact_1986} have shown that it belongs to the universality class of $\text{SU}(N){1}$ Wess-Zumino-Witten (WZW) models with central charge $c = N - 1$\cite{affleck_exact_1986,knizhnik_current_1984,di_francesco_conformal_1997}.
It is known that the $\text{SU}(N)_{1}$ WZW CFT describes the stable fixed point of the generic 1D gapless system with $\text{SU}(N)$ symmetry.
As an integrable model, the $\text{SU}(N)$ Heisenberg chain was solved by Bethe Ansatz\cite{sutherland_model_1975}.
It has also been extensively studied numerically via exact diagonalization (ED)\cite{nataf_exact_2016,wan_exact_2017},  variational Monte Carlo\cite{vörös_2021}, and DMRG methods\cite{fuhringer_dmrg_2008,nataf_density_2018,fromholz_2019,gozel_2020,nataf_edge_2021}.

In this paper, we determine the phase diagram of the $\text{SU}(3)$ \emph{t-J} chain by analyzing the emergence of a spin gap and the divergence of compressibility.
The correlation functions and structure factors are studied numerically to characterize each phase.
Comparisons are drawn with the $\text{SU}(2)$ case \cite{moreno_ground-state_2011} and the 1D $\text{SU}(N)$ Fermi-Hubbard model \cite{assaraf_metal-insulator_1999,capponi_phases_2016} to identify the distinct phases.
We identify a quasi-long-range molecular superfluid (MS) order involving three fermions on different, partially connected sites. The region occupied by the MS order phase in the phase diagram is notably smaller than the gapped superconducting region in the $\text{SU}(2)$ case. This compression is attributed to reduced clustering energy gain and enhanced fluctuations, stemming from the larger phase space of the singlet states.

Our paper is organized as follows.
In Sec. \ref{Model}, we introduce the 1D $\text{SU}(3)$ \emph{t-J} model and the setting of DMRG simulation.
In Sec. \ref{PD}, we present the phase diagram and the identification of the phase boundary.
In Sec. \ref{CF}, we calculate the structure factor and compare it with its $\text{SU}(2)$ counterpart.
In Sec. \ref{3eLEL}, the Luther-Emery liquid phase is further examined by calculating energy response and correlation functions.
The summary and discussion are in the final Sec. \ref{SD}.

\section{\label{Model}Model and Method}

In theoretical studies, $\text{SU}(N)$ Hubbard model is usually mentioned to describe the alkaline earth atoms trapped in an optical lattice.
The $\text{SU}(N)$ generalization of the Fermi-Hubbard model contains nearest-neighbor hopping and on-site interaction
\begin{equation}\label{eq1}
H_{\text{Hubbard}}^{\text{SU}(N)}=-t \sum_{\left<i,j\right>,\alpha}\left({c _{i\alpha}^{\dagger}c_{j\alpha}}+\text{H.c.}\right)+\frac{U}{2}\sum_{i}{ n_{i} \left( n_{i}-1 \right)},
\end{equation}
where $i,j$ labels chain sites, $\alpha = 1,\ldots,N$ is the $\text{SU}(N)$ nuclear spin index.
This model is invariant under the global $U\left(1\right)$ symmetry, ensuring the conservation of the total atom number.
Moreover, the $\text{SU}(N)$ symmetry arises: $c_{\alpha,i}\rightarrow \sum_{\beta}U_{\alpha\beta} c_{\beta,i}$, $U$ being an $\text{SU}(N)$ matrix.
The actual symmetry group of the Fermi-Hubbard Hamiltonian is $U(N)=U(1)\times\text{SU}(N)$.

For the large $U$ limit, the on-site repulsion confines the local state to two possible particle number occupation states around the average filling $n$: $\left[n\right]$ and $\left[n\right]+1$.
In the same manner as the $\text{SU}(2)$ case, the 1D $\text{SU}(N)$ \emph{t-J} model can be derived in second-order perturbation theory by projecting out the doubly occupied states\cite{he_2022}
\begin{equation}\label{eq2}
\begin{aligned}
H_{t-J}^{\text{SU}(N)}=-t&\sum_{\left<i,j\right>,\alpha}\mathcal P\left({c _{i\alpha}^{\dagger}c_{j\alpha}}+\text{H.c.}\right)\mathcal{P}\\\
&+J\sum_{\left<i,j\right>,\lambda}\left(\bm{T}_{i}^{\lambda\dagger}\bm{T}_{j}^{\lambda}-\gamma n_i n_j\right),
\end{aligned}
\end{equation}
where $\bm{T}_{i}^{\lambda\dagger}$ and $\bm{T}_{i}^{\lambda}$ with $\lambda=1,\ldots ,N^{2}-1$ denote the $\text{SU}(N)$ spin operators and their Hermite conjugates.
They get their explicit expression from the $\text{SU}(N)$ Lie algebra generators.
In the fermion case, the local state space is described by those antisymmetric representations of the $\text{SU}(N)$ group.
For instance, when the average local density $n$ satisfies $n < 1$, the local degree of freedom is described by the direct sum of the trivial hole $\left(singlet\right)$ and the N-dimensional fundamental, irreducible representation of $\text{SU}(N)$ $\left(\Box\right)$.
The coefficient $\gamma$ in the interaction term depends on the choice of generator basis.
We choose the interaction Hamiltonian in a simple form with zero trace
\begin{equation}\label{eq3}
\begin{aligned}
H_{t-J}^{\textrm{SU}(N)}=-t&\sum_{\left<i,j\right>,\alpha}\mathcal{P}\left({c _{i\alpha}^{\dagger}c_{j\alpha}}+\text{H.c.}\right)\mathcal{P}\\\
&+J\sum_{\left<i,j\right>,\alpha,\beta}\left(c_{i\alpha}^{\dagger}c_{i\beta}c_{j\beta}^{\dagger}c_{j\alpha}-n_{i\alpha}n_{j\beta}\right),
\end{aligned}
\end{equation}
where the interaction term is consistent with the usual form
\begin{equation}\label{eq4}
H^{\text{SU}(N)}_J=J\sum_{i,j} \frac{1}{2}\left(P_{ij}-\frac{1}{N}\right),
\end{equation}
with 2-site permutation operator $P_{ij}$ exchanges the states on sites $i$ and $j$.

We use the DMRG algorithm to study $\text{SU}(3)$ \emph{t-J} chain with open boundary conditions and the length up to $L=120$ sites.
The DMRG algorithm conserves the total particle number $N_{tot}=nL$, which is set to multiples of 3 to ensure color neutrality.
We extrapolate the thermodynamic limit by studying systems of lengths $L=30, 60, and 90$.
These calculations use the ITensor library\cite{fishman_itensor_2020} DMRG update, with a bond dimension of up to $m=5,000$.
The maximum truncation error of ground state DMRG simulation is on the order of $10^{-8}$.
At least 100 sweeps are performed for all density and interaction parameters.

We calculated the energy gap to $\text{SU}(3)$ spin-flip excitations to characterize the phases and determine their boundaries.
Various correlation functions were also calculated to characterize the distinct phases, such as the density-density correlation function,
\begin{equation}\label{eq5}
N_{ij} = \langle n_i n_j\rangle - \langle n_i\rangle \langle n_j\rangle.
\end{equation}
The $\text{SU}(3)$ spin-spin correlation function is calculated as
\begin{equation}\label{eq6}
T_{ij} = \sum_{\lambda=1}^{N^{2}-1} (\langle \bm{T}^\lambda_i \bm{T}^\lambda_j\rangle -\langle \bm{T}^\lambda_i\rangle \langle \bm{T}^\lambda_j\rangle),
\end{equation}
where $\bm{T}^\lambda$ are chosen to be the Hermite generator of the fundamental representation.
The explicit forms of $T^3$ and $T^8$ are given as
\begin{equation}\label{eq7}
T^3 = \frac{1}{2} \begin{pmatrix} 1 & 0 & 0 \\ 0 & -1 & 0\\ 0 & 0 & 0 \end{pmatrix},
T^8 = \frac{1}{2\sqrt{3}} \begin{pmatrix} 1 & 0 & 0 \\ 0 & 1 & 0\\ 0 & 0 & -2 \end{pmatrix}.
\end{equation}

These two $\text{SU}(3)$ spin components are the two diagonal matrices of the Gell-Mann matrices. They are orthogonal to the density channel.
The corresponding structure factors of a certain two-point correlation $X_{ij}$ are obtained by a Fourier transform
\begin{equation}\label{eq9}
X (k) = \frac{1}{L}\sum_{i,j=1}^{L}{e^{ik\left(x_i -x_j \right)}X_{ij}}.
\end{equation}
Even though the system we calculated lacks translational invariance for the open boundary conditions, the effect of finite size could hardly be observed in the Fourier transformation procedure.

\section{\label{PD}Phase Diagram}

\begin{figure}[t]
	\includegraphics[width=0.99\linewidth]{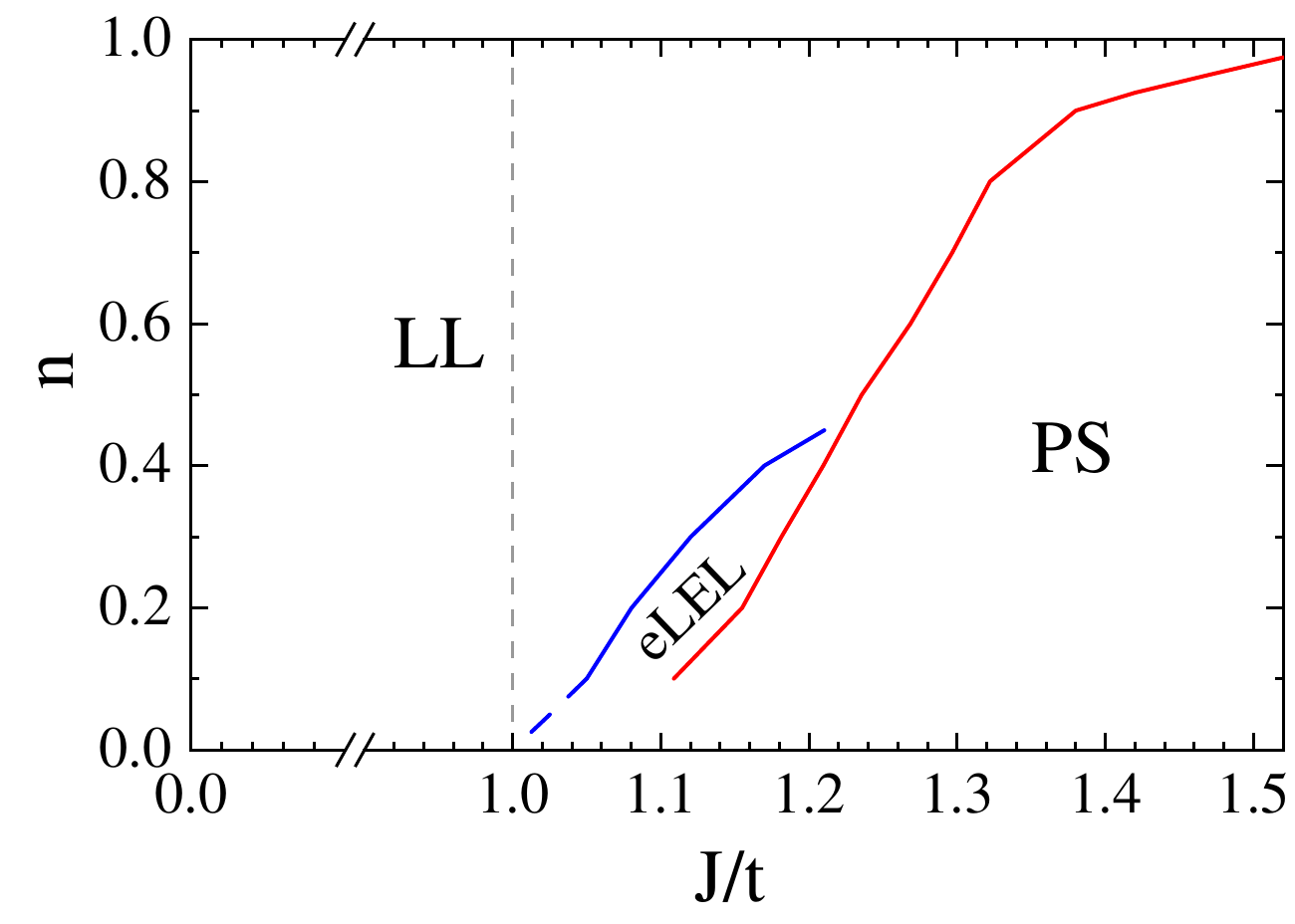}
	\centering
	\caption{Phase diagram of the 1D $\text{SU}(3)$ \emph{t-J} model from DMRG for densities $0.1\le n \le 0.95$ and in the range $0<J/t\le 1.5$, where $n=N/L$ is the particle density. The grey dashed line $J=t$ is the exactly solvable parameter regime. }
	\label{phasediagram}
\end{figure}

The phase diagram for the $\text{SU}(3)$ \emph{t-J} model (FIG. \ref{phasediagram}.) contains three phases: Luttinger liquid (LL), a spin-gapped phase named extended Luther-Emery liquid (LEL), and phase separation (PS).
It differs from the phase diagram of $\text{SU}(2)$ case mainly in the position of the phase boundary.

\subsection{\label{LL} Luttinger Liquid}
We can characterize this phase through the value of the Luttinger parameter $K_{\rho}$, with $K_{\rho}<1$ for a repulsive interaction on the density sector, $K_{\rho}=1$ and $K_{\rho}>1$ for the free and attractive case.
We obtain the structure factor for the density correlation to compute $K_{\rho}$.
$N(k)$ does not show qualitative difference from the $\text{SU}(2)$ case.
They both show a linear behavior with a slope proportional to $K_{\rho}$ in $k\rightarrow0$ limit\cite{giamarchi_quantum_2004,clay_possible_1999,ejima_tomonaga-luttinger_2005, Manmana_2011}
\begin{equation}\label{eq10}
N\left(k\rightarrow 0\right)\rightarrow \frac{NK_{\rho}\left|k\right|a}{2\pi}.
\end{equation}
where $a$ is the lattice constant.
It can be derived from the Fourier transformation of the inverse quadratic term of the asymptotic density correlation\cite{Manmana_2011}.
\begin{equation}\label{eqNr}
  N_{r,0}\sim n^2-\frac{NK_{\rho}}{2\left(\pi r\right)^2}+A_1 \frac{\cos \left(2k_{F}r\right)}{r^{2K_{\rho}/N+2-2/N}}.
\end{equation}
By calculating density structure factors for different $n$ and $J$, we find that $K_{\rho} \rightarrow 1/3$ when $J \rightarrow 0$ for any densities below 1, which results from the same slope 0.5 as in $\text{SU}(2)$ case.
The physical property of density, spin correlations, and corresponding structure factors is further discussed in Sec. \ref{SF}.

\subsection{\label{SG}Spin-Gap Phase}
As $J$ increases, the value of $K_{\rho}$ grows more significant than 1, and the system gets into the attractive phase.
Meanwhile, a spin gap appears in a low-density region.

\begin{figure}[t]
	\includegraphics[width=0.99\linewidth]{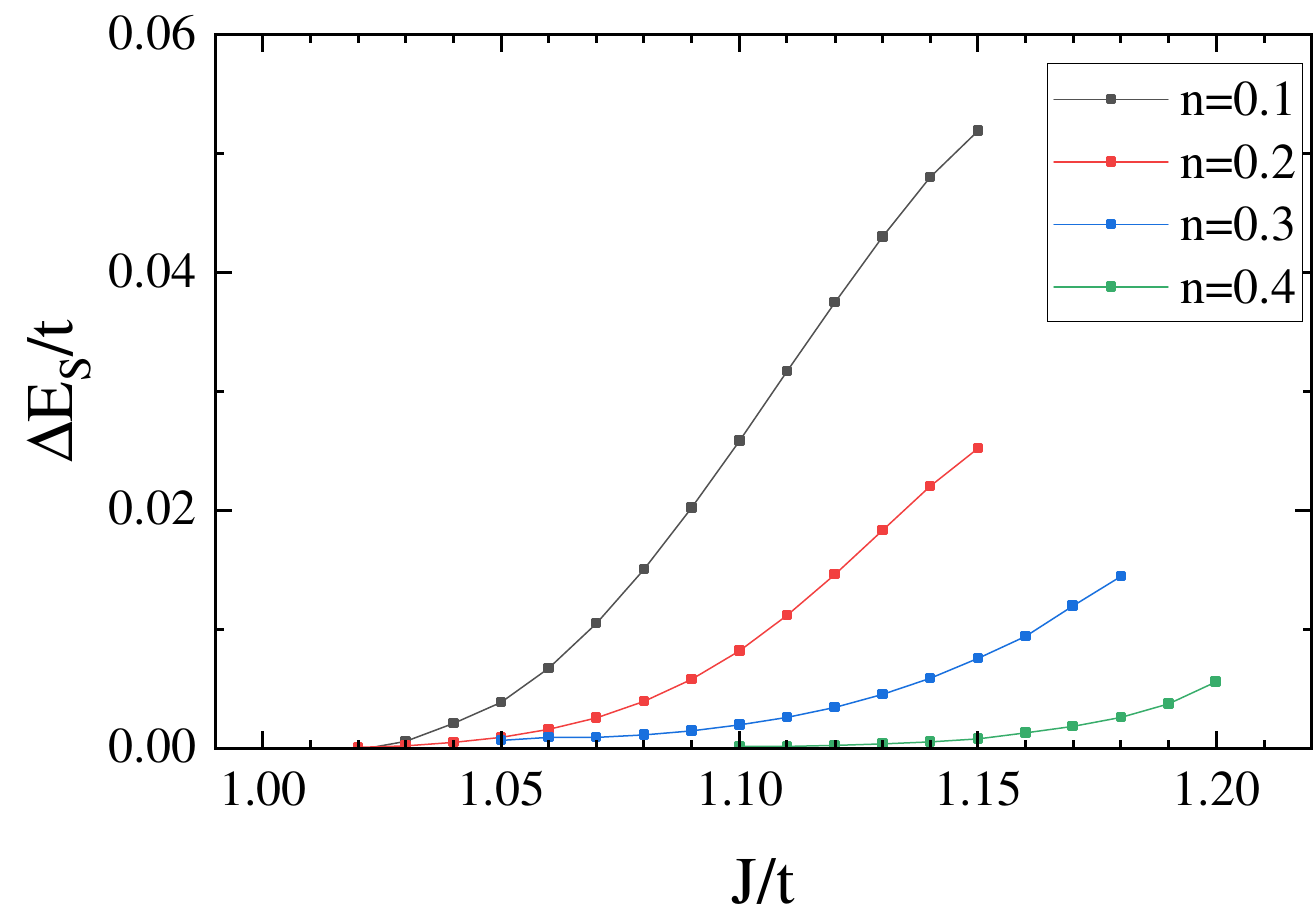}
	\centering
	\caption{Spin gap $\Delta E_S$ in the thermodynamic limit as a function of $J/t$ in the range $1.0-1.2$ for different densities $n=0.1$, $0.2$, $0.3$, and $0.4$. The values of $\Delta E_S$ are extrapolated from finite-size scaling using system sizes $L=30, 60, 90$, and $120$. }
	\label{spingap}
\end{figure}
The spin gap $\Delta E_{S}$ is measured directly to find the spin gap region.
It is defined as the spin excitation energy from a singlet to a spin-flipped state, which is the energy difference
\begin{equation}\label{eq11}
\begin{aligned}
\Delta E_{S}=&E_0 \left(N, T^3_{tot} = 1, T^8_{tot} = 0\right) \\
&- E_0 \left(N, T^3_{tot} = 0, T^8_{tot} = 0\right),
\end{aligned}
\end{equation}
where the subscript $0$ means the lowest-energy level with given quantum numbers $N$, $T^3_{tot}$, and $T^8_{tot}$.
We use second-order polynomial fits to extrapolate the spin gap from finite system sizes ($L = 30, 60, 90,$ and $120$) to the thermodynamic limit, based on the spin excitation energy versus $1/L$ for $n = 0.1$ to $0.4$.
The extrapolated results are presented in Fig. \ref{spingap}, where the horizontal axis corresponds to $J/t$ in the range $1.0$ to $1.2$. 
The results show that the spin gap increases with $J/t$, starting from near zero at $J/t \leq 1.0$ and becoming finite at larger values of $J/t$, with the growth more pronounced at lower densities.

This calculation is performed for different densities $n$ and $J$ to obtain the region with a spin gap in FIG. \ref{spingap}.
A finite spin gap is observed when $J$ increases more remarkably than the supersymmetric value $J = 1.0t$.
We define the phase boundary in the phase diagram (FIG. \ref{phasediagram}) as the value of $J$ for which $\Delta E_S >10^{-3}t$ to avoid the effects of numerical and fitting errors.
The spin gap implies the existence of the regime in which pairs or trions form.
This phase, characterized by a quasi-long-range molecular superfluid order formed by three particles, is identified as the extended LEL phase.
A more detailed discussion of the clustering mechanics is addressed in Sec. \ref{3eLEL}.

\subsection{\label{PS}Phase Separation}
As the interaction $J$ grows, the attraction among the particles becomes strong enough that the particles start to form $\text{SU}(3)$ antiferromagnetic domains.
The system is separated into particle-rich and hole-rich regions.
Similar to the $\text{SU}(2)$ case, all particles condensate in a single island in $J \rightarrow \infty$ limit.
This phase, referred to as the electron solid phase, was proposed by Chen and Moukouri\cite{chen_numerical_1996}.
In this phase, the kinetic term is heavily suppressed, and spin fluctuations are described by the $\text{SU}(3)$ Heisenberg model.
The divergence of compressibility is used to detect this phase, with its inverse vanishing at the threshold."

\begin{figure}[t]
	\includegraphics[width=0.99\linewidth]{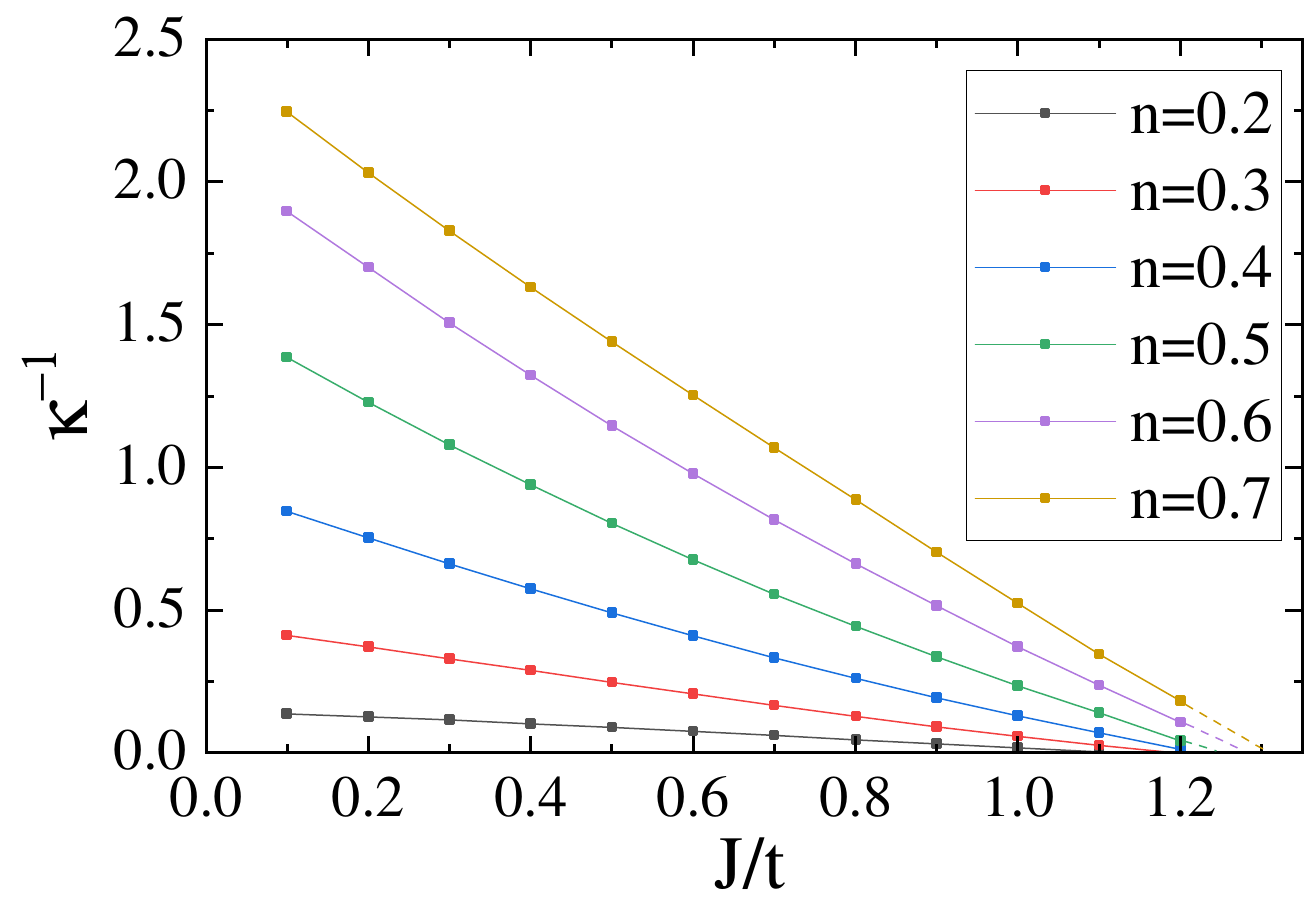}
	\centering
	\caption{Inverse of the compressibility $\kappa^{-1}$ as a function of $J$ for different density.  }
	\label{kappa}
\end{figure}

The expression for the ground state inverse compressibility is given by
\begin{equation}\label{eq12}
\begin{aligned}
\kappa ^{-1}(n) =& n^2 \frac{\partial ^2 e_0 (n)}{\partial n^2} \\
\approx & n^2\frac{\left[e(n+\Delta n)+e(n- \Delta n)-2e(n)\right]}{\Delta n^2},
\end{aligned}
\end{equation}
where $e_0\left(n\right) = E_0 /L$ is the energy per site, and the second line shows the approximation for practical numerical calculation when the change in the density is finite.
FIG. \ref{kappa} shows $\kappa^{-1}$ versus $J/t$ for different densities.
We can locate the boundary of the phase-separated phase by finding the critical value $J_c\left(n\right)$ where $\kappa^{-1}$ vanishes.

\section{\label{CF}Correlation Functions}

\subsection{\label{SF}Structure Factors}
\begin{figure}[t]
	\includegraphics[width=0.99\linewidth]{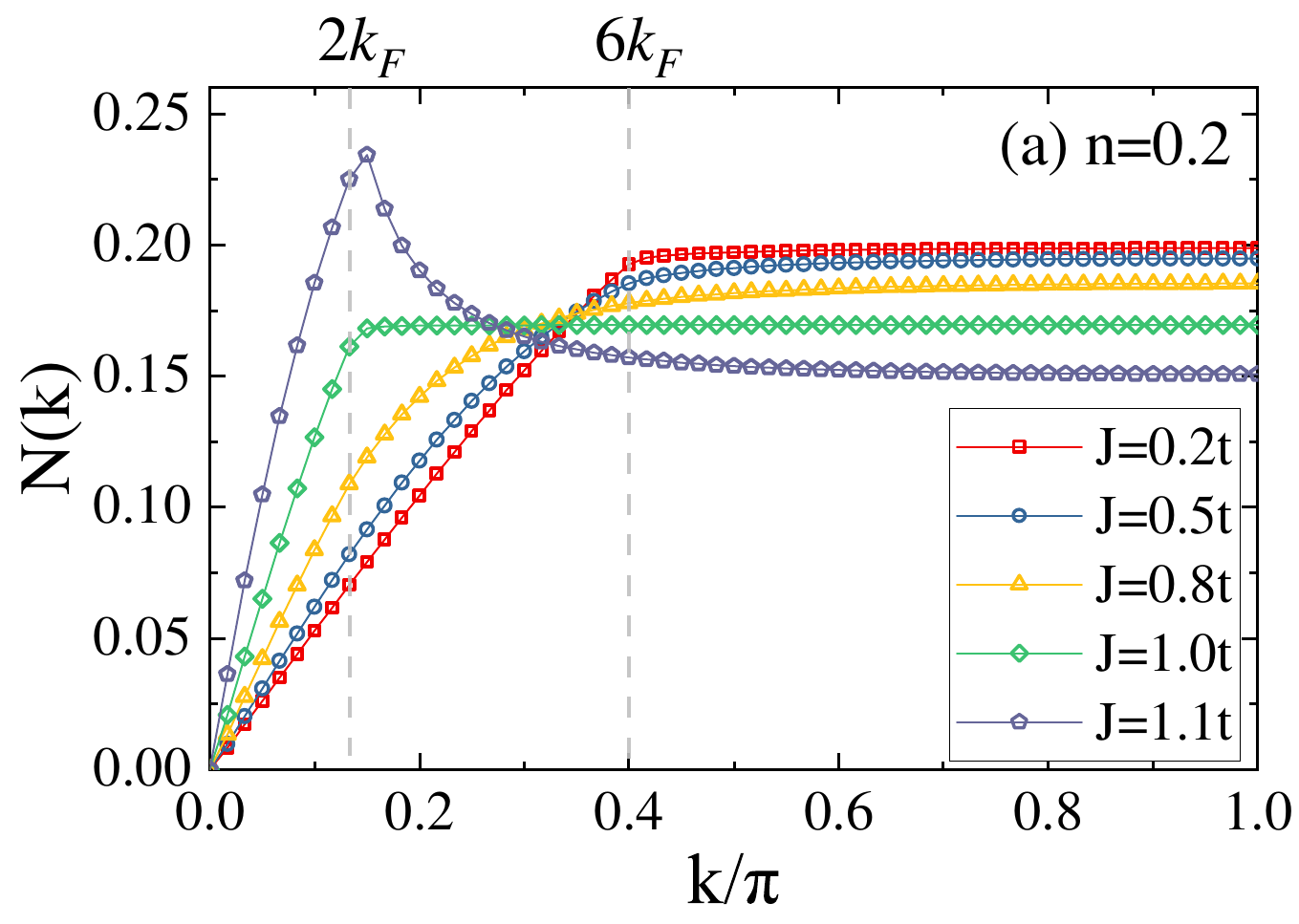}
    \includegraphics[width=0.99\linewidth]{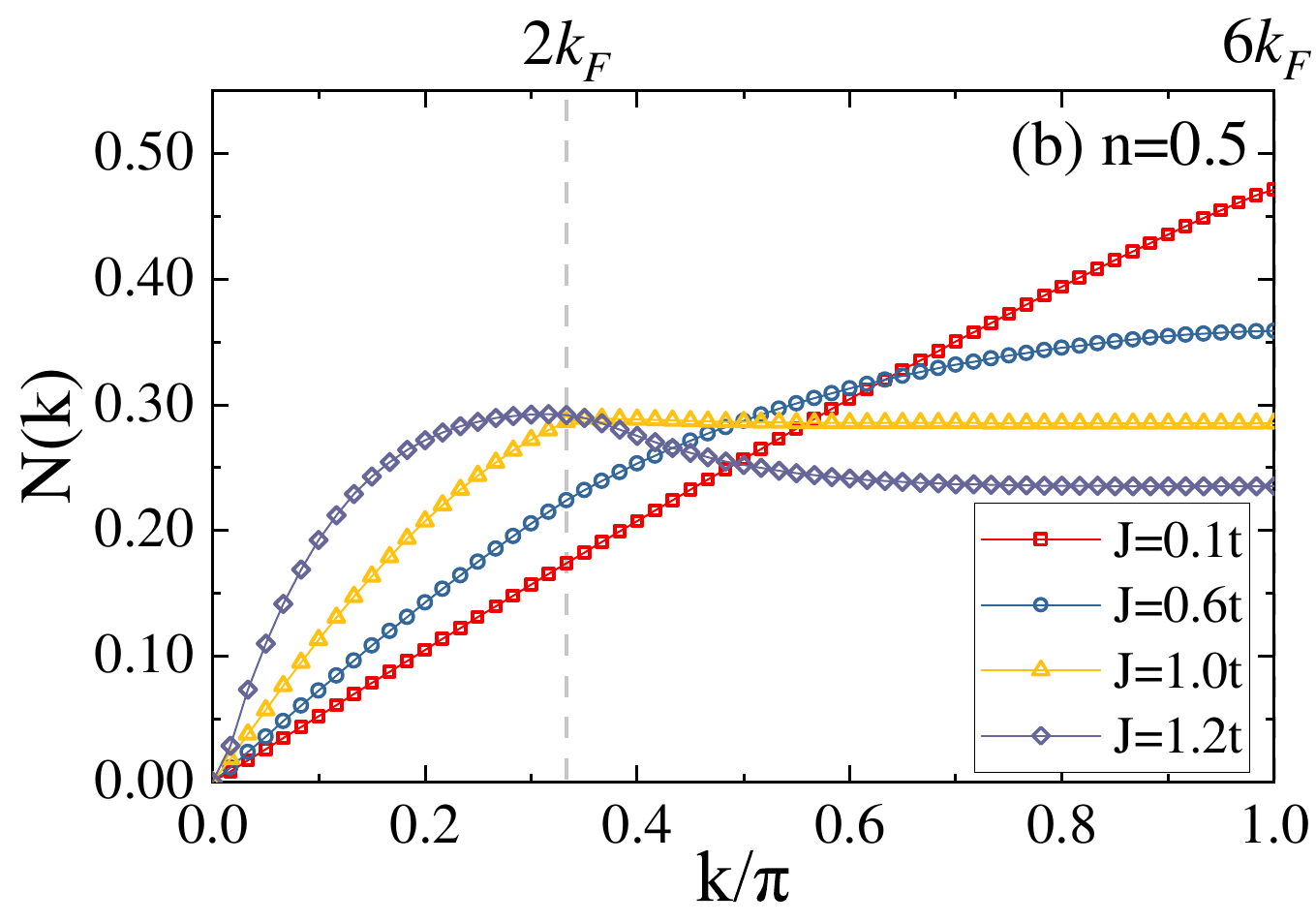}
    \includegraphics[width=0.99\linewidth]{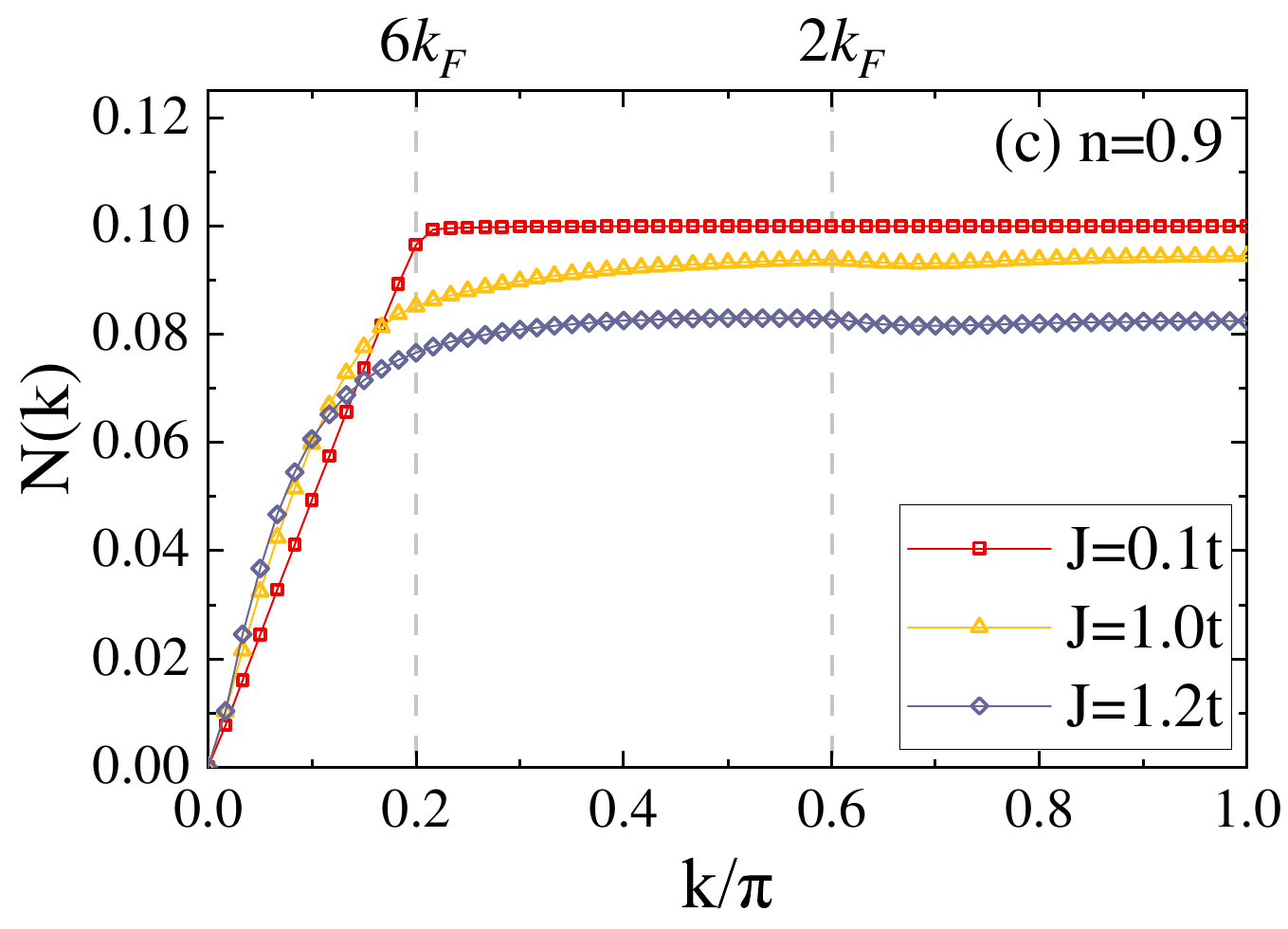}
	\centering
	\caption{Structure factor $N(k)$ for density-density correlation function for $L=120$ and for different values of $n$ and $J$. The location of $6k_F$ is determined by folding them back to the outer sector in the first Brillouin zone. }
	\label{Nk}
\end{figure}

\begin{figure}[t]
	\includegraphics[width=0.99\linewidth]{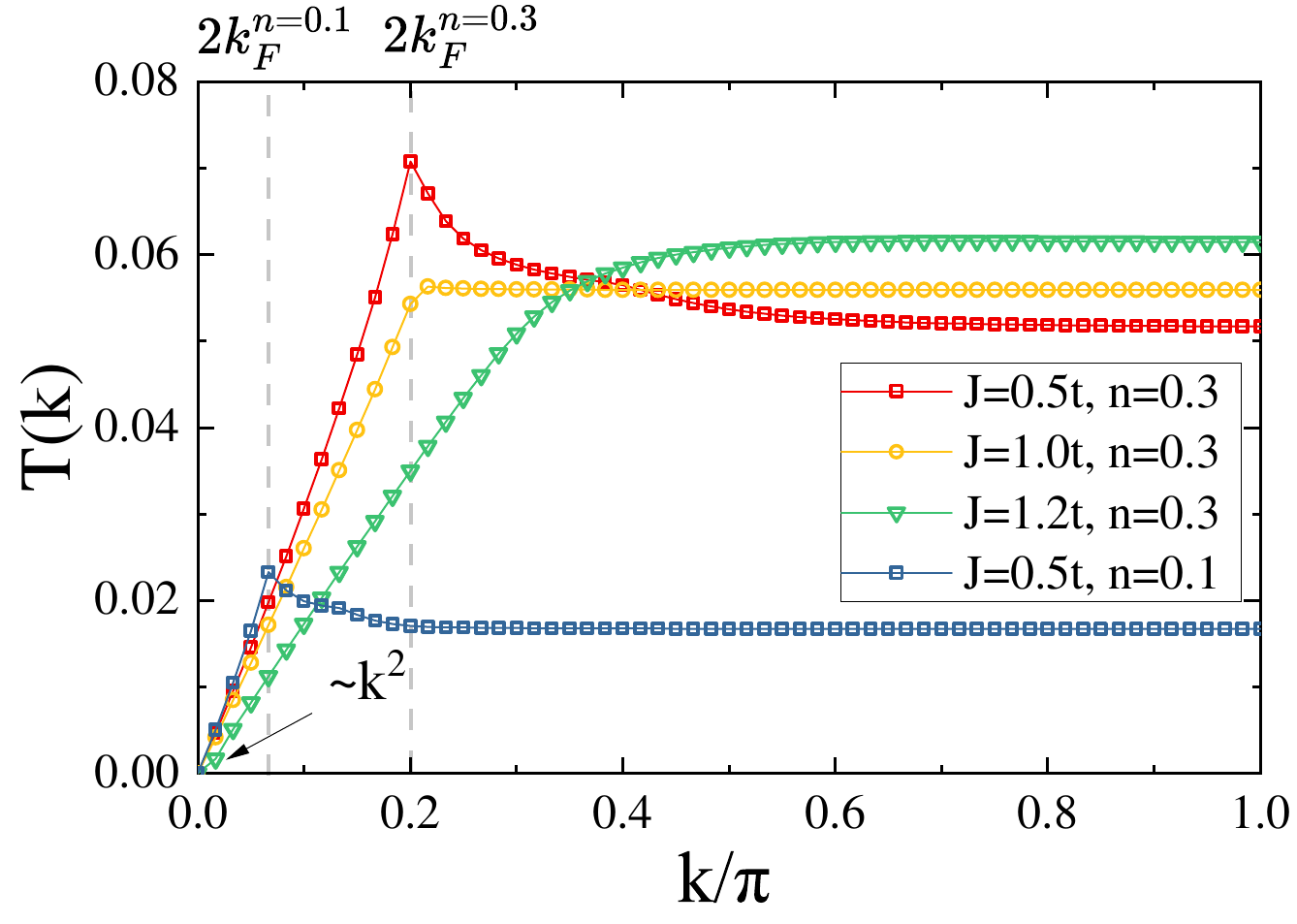}
	\centering
	\caption{ Structure factor $T(k)$ for $\text{SU}(3)$ spin-spin correlation function for $L=120$ and for different values of $n$ and $J$. }
	\label{Tk1}
\end{figure}

The phase diagram of the 1D $\text{SU}(3)$ \emph{t-J} model appears distorted compared to the $\text{SU}(2)$ case.
However, significant deviations are observed in correlation functions.
The structure factors for certain correlation functions provide a more detailed characterization of the different phases.
They are displayed in FIG. \ref{Nk} and \ref{Tk1}.

FIG. \ref{Nk} shows the structure factor $N(k)$ for the density-density correlation function.
We plot several values of $J$ for $n=0.2, 0.5$, and $0.9$.
We can see $N(k=0)=0$ due to the conversation of total particle numbers for all cases.
Similarly, linear behavior is observed as $k\rightarrow 0$, indicating the gapless nature of the charge sector.

For $J\ll t$, the ground state is a repulsive LL phase.
An anomaly at $6k_F$ $\left(k^{\text{SU}(3)}_{F} = n\pi /3\right)$ can be observed, which is different from the typical anomaly at $4k_F$ $\left(k^{\text{SU}(2)}_{F} = n\pi /2\right)$ in the $\text{SU}(2)$ \emph{t-J} model and cannot be reflected in Eq.\ref{eqSr}.
The two anomalies are located at wave vectors with the same form when we express them in the manner of density $n$ $\left(6k^{\text{SU}(3)}_{F} = 4k^{\text{SU}(2)}_{F} = 2n\pi\right)$.
They correspond precisely to a charge modulation with a spatial distance periodicity between the electrons.
Due to repulsive interactions between electrons, they tend to form a 1D analog of a Wigner crystal\cite{giamarchi_quantum_2004,wigner_interaction_1934,mahan_many-particle_2000}.
In 1D, this manifests as a charge-ordering pattern, but due to strong quantum fluctuation, it lacks even quasi-long-range order.
We expect a slight anomaly at \(2Nk_F\) in the density structure factor of the 1D $\text{SU}(N)$ \emph{t-J} model.
This is different from the one in the $\text{SU}(N)$ Fermi-Hubbard model because double occupancy is forbidden in the \emph{t-J} model.

Like the $\text{SU}(2)$ case, as $J$ increases, the $6k_F$ anomaly is suppressed, and a $2k_F$ peak is formed.
This signifies a $2k_F$ charge density wave, corresponding to the pairing charge density wave with $2k_F$ in $\text{SU}(2)$ case.
However, considering the wavelength expressed in terms of density,
\begin{equation}\label{eq13}
\lambda = \frac{2\pi L}{2k_F} = \frac{3L}{n},
\end{equation}
it reflects a distance three times the lattice spacing of individual particles, signaling 3-fermion pairing within a wavelet.
In the attracting regime, fermions combine in a different way compared to the $\text{SU}(2)$ pairing case.
This supports that the singlet formation of 3 spins still dominates the attractive zone despite the less connection between spins in the \emph{t-J} model.
By comparing different densities, we observe that the $2k_F$ apex is suppressed as particle density increases, which signals weaker 3-clustering in the density channel for higher density.
The peaks and anomalies are slightly more prominent than $2k_F$ and $6k_F$ in the $N(k)$ diagram.
It results from the open boundary condition of our DMRG simulation.
The local density is suppressed near the boundaries, so the effective density is larger than the total charge and system size ratio.

In FIG. \ref{Tk1}, the structure factor $T(k)$ for the spin-spin correlation function is shown.
For $J<t$ at all densities, the tendency to antiferromagnetism is revealed by a peak at $2k_F$.
It results from the quasi-long-range $\text{SU}(3)$ antiferromagnetic order.
The slopes near the $k\rightarrow 0$ limit remain almost unchanged in the repulsive region.
It is consistent with the quadratic term in the asymptotic spin correlation function in repulsive Luttinger liquid.
\begin{equation}\label{eqSr}
  T_{r,0}\sim -\frac{1}{2\left(\pi r\right)^2}+B_1 \frac{\cos \left(2k_{F}r\right)}{r^{2K_{\rho}/N+2-2/N}}.
\end{equation}
As $J$ increases, the peak is suppressed, implying the existence of a molecular superfluid phase.
The spin gap develops in low density, signaling a phase like the LEL phase with the singularity at $2k_F$ completely suppressed.
This covers the MS phase regimes.
It shares some behaviors with the typical LEL phase, such as exponential decay of the spin-spin correlation function and the quadratic behavior at small $k^{\prime}$s in the structure factor.

\section{\label{3eLEL}Trion and Extended LEL Phase}
\begin{figure}[htbp]
	\includegraphics[width=0.99\linewidth]{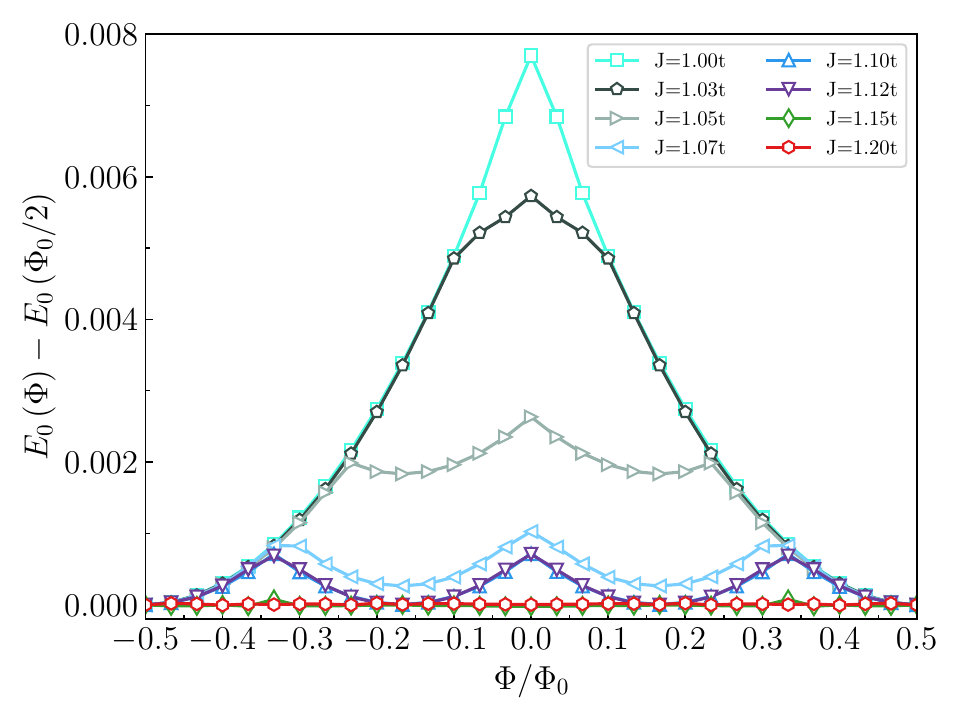}
	\centering
	\caption{Ground state energy $E_0\left(\Phi\right)$ with respect to $E_0\left(\Phi_0 / 2\right)$ for $\text{SU}(3)$ \emph{t-J} flux model(Eq. \ref{eq14}).
The data of negative $\Phi$ is copied from its positive counterpart for the parity symmetry.
}\label{Ephi}
\end{figure}

The peaks of FIG. \ref{Nk} at $2k_F$ and the suppressed anomaly in FIG. \ref{Tk1} implies the existence of a 3-clustering phase.
$\text{SU}(3)$ particles in fundamental representations cluster into singlets as $J$ increases.
This clustering is expected to be weaker than pairing because the interaction forming a singlet cannot directly link three $\text{SU}(3)$ particles.
The strong fluctuation in the 1D system may break the long-range clustering order with higher translational symmetry breaking.
For the \emph{t-J} model, A. Seidel \emph {et~al.} proved that the energy spin-gapped $\text{SU}(2)$-invariant Luttinger liquids have $hc/2e$ flux periodicity for the broken translational symmetry with a doubling of the unit cell\cite{seidel_flux_2004,seidel_luther-emery_2005}.
To reveal the 3-clustering state, we calculate the energy dependence on the flux in the periodic boundary condition
\begin{equation}\label{eq14}
\begin{aligned}
H \left(\Phi\right)=& -t\sum_{\left<i,j\right>,\alpha}\mathcal{P}\left(e^{\left(2\pi i /L\right) \Phi/ \Phi_0}{c _{i\alpha}^{\dagger}c_{j\alpha}}+H.c.\right)\mathcal{P}+H_J,
\end{aligned}
\end{equation}
in spin-gap opening region($n=0.1,J=1.00-1.20t$).
$\Phi_0\equiv hc/e$ is the elementary flux quantum.
The results are shown in FIG. \ref{Ephi}.
We observe a branch of quadratic ground state energy spectrum declining as $J$ increases, referring to $E_0\left(\Phi = \Phi_0 / 2\right)$.
The gap between bottoms of quadratic spectra will close when J exceeds $1.10t$, which leads to a $\Phi_0 /3$ periodicity.
By analogy, this indicates the combination of trion carriers with the $\Phi_0 /2$ periodicity for the superconducting phase in the $\text{SU}(2)$ case.


This periodicity has also been investigated by W.J. Chetcuti \emph {et~al.} \cite{chetcuti_persistent_2022,chetcuti_probe_2023}.
They observe the same behavior from small $U$ to large $U$ in $\text{SU}(3)$ Fermi-Hubbard model under flux penetration through the ring for incommensurate fillings.
This demonstrates the paramagnetic persistent current behavior, which supports the 3-clustering mechanism at the mesoscopic scale.

The N-clustering feature aligns with the characteristics of the MS phase in 1D $\text{SU}(N)$ Fermi-Hubbard model with $N>2$ \cite{capponi_phases_2016}.
In the $U<0$ region, the LEL phase is discovered at a sufficiently low density.
The superfluid instability is a molecular one with the order parameter $M_i=c^{\dagger}_{i,1}c^{\dagger}_{i,2}\cdot \cdot \cdot c^{\dagger}_{i, N}$.
This corresponds to the formation of $\text{SU}(N)$ singlet by $N$ fermions on a single site.
This phase with $U<0$ is governed by the competition between the CDW and molecular superfluid (MS) orders.

In $\text{SU}(3)$ \emph{t-J} model, the local singlet is forbidden by the occupation constraint, and thus the clustering order is defined on three adjacent sites
\begin{equation}\label{eq15}
C(i)=\frac{1}{\sqrt{6}}\sum^{3}_{\alpha,\beta,\gamma=1}\varepsilon^{\alpha\beta\gamma}c_{i,\alpha}c_{i+1,\beta}c_{i+2,\gamma},
\end{equation}
where $\varepsilon^{\alpha\beta\gamma}$ is the antisymmetric tensor. We compare the correlation function for the clustering order
\begin{equation}\label{eq16}
C(x)=\left<C(i)C^{\dagger}(i+x)\right>,
\end{equation}
with the correlation of pairing order,
\begin{equation}\label{eq17}
\Delta^{\alpha}(x)=\left<\Delta^{\alpha}(i)\Delta^{\alpha\dagger}(i+x)\right>,
\end{equation}
where $\Delta^{\alpha}(i)=\sum_{\beta\gamma}\varepsilon^{\alpha\beta\gamma}c_{i,\beta}c_{i+1,\gamma}$, and CDW correlation
\begin{equation}\label{eq19}
N(x)=\left<N_{i,i+x}\right>.
\end{equation}

\begin{figure}[bth]
	\includegraphics[width=0.48\linewidth]{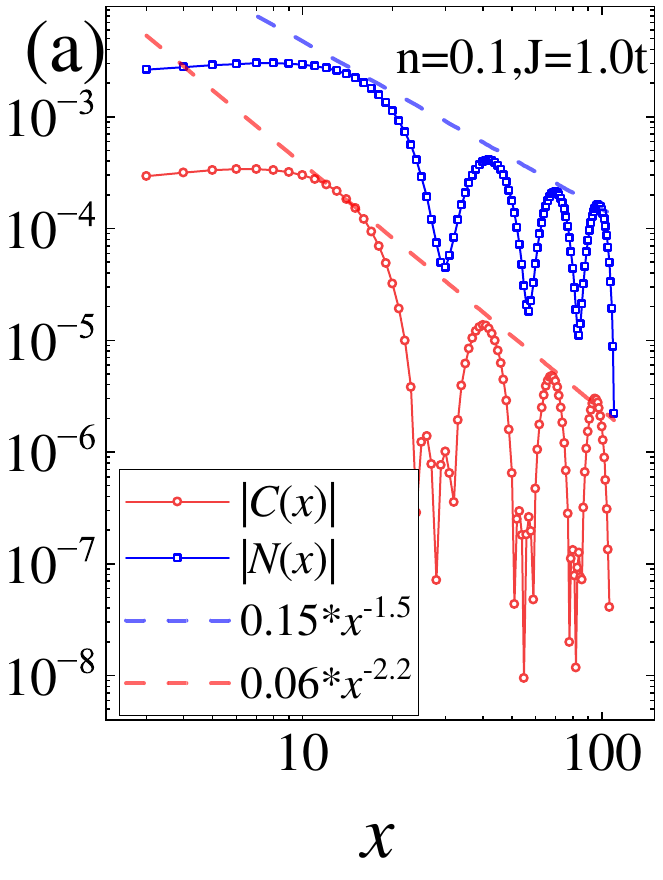}
  \includegraphics[width=0.48\linewidth]{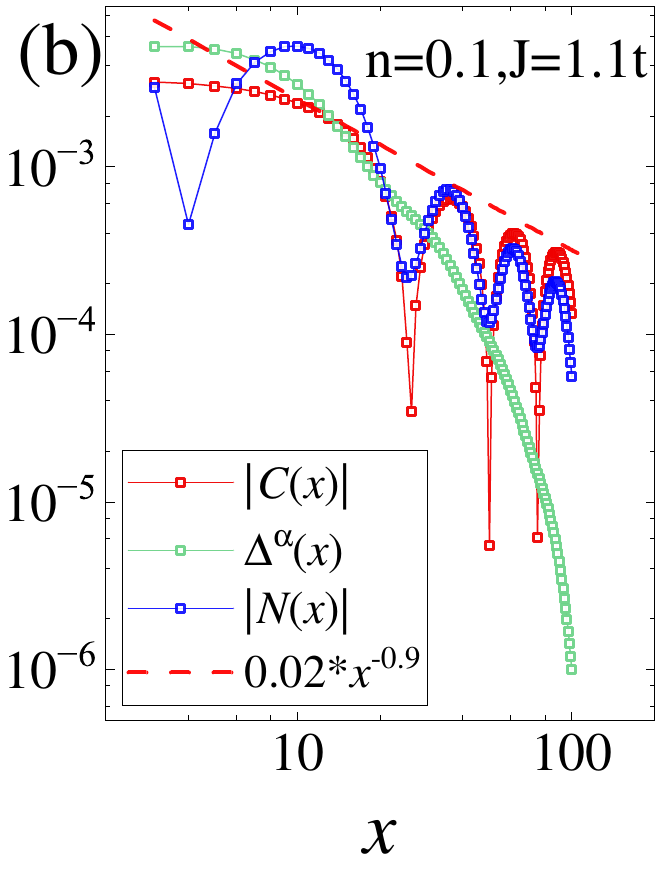}
	\centering
	\caption{MS $\left|C(x)\right|$, pairing $\Delta^{\alpha}(x)$ and density $N(x)$ correlation functions in LL (a) and LEL (b) phase. The upper envelop line shows the algebraic decay of corresponding correlation functions.}\label{fig_MSorder}
\end{figure}

Fig. \ref{fig_MSorder}(a) shows the decay of density-density and MS correlation in real space.
The upper envelope curves of both correlation functions suggest that they decay as the power law when the wave part is removed.
The lower decaying rate shows that the CDW quasi-long-range order dominates in the LL regime.
FIG. \ref{fig_MSorder}(b) shows the behavior of the three correlation functions for the LEL phase.
We can observe that the envelope curve of the singlet clustering correlation function shows the power law decay behavior.
The 3-clustering order decays slower than the pairing and CDW orders, regarded as the dominant order.
The asymptotic form of the correlation functions is 
\begin{equation}\label{powerlaw_nr}
  N(x) \sim \cos\left(2k_F x\right)x^{-\alpha_N},
\end{equation}
\begin{equation}\label{powerlaw_c6}
  C(x) \sim \sin\left(k_F x\right)x^{-\alpha_C}.
\end{equation}
Comparing the two correlation function diagrams, we find that the MS order becomes more dominant when the spin gap opens than the CDW order.
Given the dominant MS order, the $hc/3e$ flux periodicity, and the finite spin gap, we identify this phase as the extended LEL phase.

\begin{figure}[htbp]
	\includegraphics[width=0.99\linewidth]{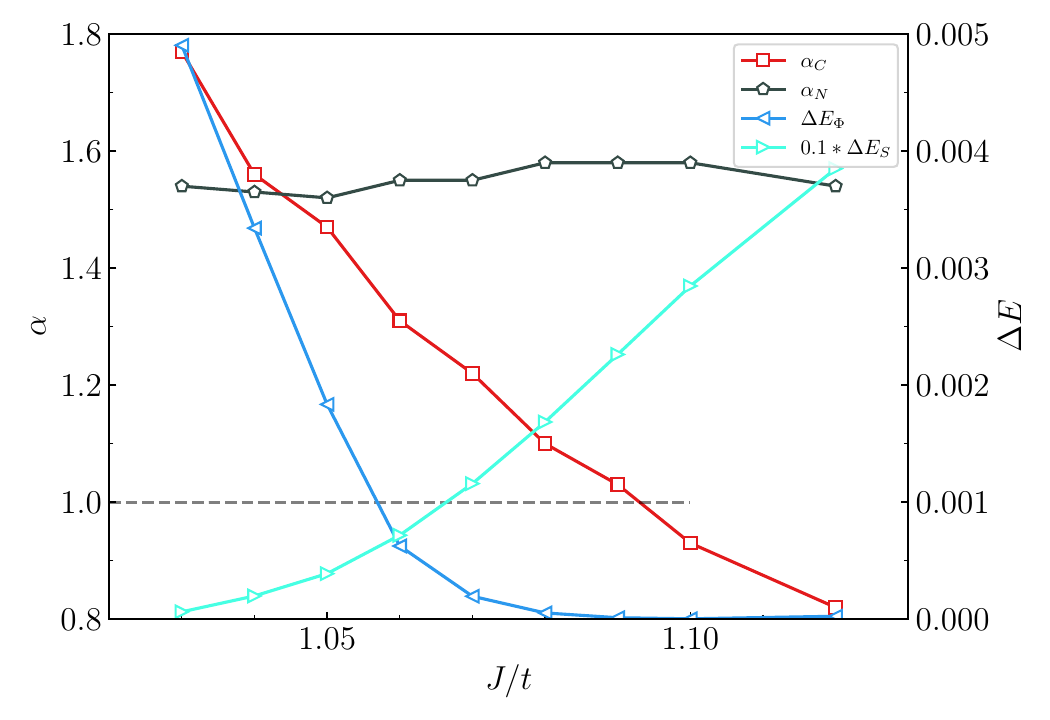}
	\centering
	\caption{%
3-clustering exponent $\alpha_C$, density correlation $\alpha_N$, energy difference between two quadratic branches $\Delta E_\Phi = E_0 \left(\Phi = 0\right) - E_0\left(\Phi = \Phi_0 / 3\right)$ in ring model(\ref{eq14}) and spin gap $\Delta E_S$ (multiplied by 0.1 to fit scale).%
}\label{alpha_gap}
\end{figure}

We calculate the exponent's dependence with $J$ and plot them with the energy difference between two quadratic branches $\Delta E_{\Phi}$ and the spin gap $\Delta E_{S}$ in FIG. \ref{alpha_gap}.
These variables indicate a phase transition in the range $J/t \in \left(1.05,1.10\right)$.
The power-law exponents differ significantly from those in the attractive Fermi-Hubbard model, which is $\alpha_N^{FH} = 2K_{\rho}/N$ and $\alpha_C^{FH} = \left(K_\rho +N^2/K_\rho \right) /\left(2N\right)$ (odd $N$)\cite{lecheminant_confinement_2005,lecheminant_competing_2008,capponi_molecular_2008}.
They do not satisfy this relationship quantitatively, implying the distinct features in this MS state.
Furthermore, the exponent of MS correlation declines to below 1, the minimum value of $\alpha_C^{FH}$, which means a more robust MS order than that in the Fermi-Hubbard model.
This inconsistency is more apparent in the $\text{SU}(3)$ $t$-$J$ model than in its $\text{SU}(2)$ counterpart.
No lower bound is greater than 0 in the MS component for even $N$, $\alpha_C^{FH} = N / 2K_\rho$.
The more extensive phase space occupation of singlets is proposed to extend the correlation range of the MS order.
A more precise calculation using bosonization may clarify the explicit relation to the Luttinger parameter $K_\rho$.

The above numerical results support the existence of the string-linked singlet simplex in the $\text{SU}(N)$ fermion model\cite{arovas_simplex_2008}.
This simplex has higher energy than the loop-like ones or more strongly linked ones.
It can be explained by the effect of one-dimensional systems extending the effective interaction range.

\section{\label{SD}Summary and discussion}
In this work, we calculate the phase diagram of the 1D $\text{SU}(3)$ \emph{t-J} model numerically by DMRG.
The phase boundary of three phases: Luttinger liquid, extended Luther-Emery liquid, and phase separation are demonstrated.
The Luttinger liquid phase has no charge or spin gaps. 
Although there is no charge gap in the extended Luther-Emery liquid phase, the emergence of a spin gap is deeply rooted in the combination of three SU(3) fermions on sites that are not completely connected. 
In the phase separation phase, the divergence of compressibility mainly identifies the system splitting into particle-rich and hole-rich regions.

Compared to the well-studied $\rm{SU}\left(2\right)$ model, the $4k_F$ anomaly in the density structure factor shifts to $6k_F$ in the $\rm{SU}\left(3\right)$ case, which reflects the tendency toward Wigner crystallization in the repulsive limit. In contrast, both models show a $2k_F$ peak in the spin correlation.
In the LEL phase, the $2k_F$ peak in the density structure factor persists, but pairing turns into three-particle clustering.
The phase separation behavior is largely similar between the two models.
Based on the comparison of the two models, we predict the $2k_F$ and $2Nk_F$ singularity for 1D $\text{SU}\left(N\right)$ \emph{t-J} model in the same filling regime.

We further characterize the extended LEL phase for its novel clustering structure through persistent currents and correlation functions.
It is confirmed that three fundamental irreducible representations form a singlet in line under antiferromagnetic interaction.
They differ from the on-site molecular superfluid singlet in the $\text{SU}(N)$ attractive Fermi-Hubbard model in the spatial structure.
This feature may affect ground state properties, leading to different power law exponents for density and MS correlation.

\section*{Acknowledgments}
This work is supported by the National Key Research and Development Program of China (Grant No. 2022YFA1404204) and the National Natural Science Foundation of China (Grants Nos. 11625416 and 12274086).
The authors acknowledge Beijing PARATERA Tech CO., Ltd. (\href{https://www.paratera.com/}{https://www.paratera.com/}) for providing HPC resources that have contributed to the numerical results reported in this paper.

\bibliographystyle{apsrev4-2}

\begin{thebibliography}{0}%
\makeatletter
\providecommand \@ifxundefined [1]{%
 \@ifx{#1\undefined}
}%
\providecommand \@ifnum [1]{%
 \ifnum #1\expandafter \@firstoftwo
 \else \expandafter \@secondoftwo
 \fi
}%
\providecommand \@ifx [1]{%
 \ifx #1\expandafter \@firstoftwo
 \else \expandafter \@secondoftwo
 \fi
}%
\providecommand \natexlab [1]{#1}%
\providecommand \enquote  [1]{``#1''}%
\providecommand \bibnamefont  [1]{#1}%
\providecommand \bibfnamefont [1]{#1}%
\providecommand \citenamefont [1]{#1}%
\providecommand \href@noop [0]{\@secondoftwo}%
\providecommand \href [0]{\begingroup \@sanitize@url \@href}%
\providecommand \@href[1]{\@@startlink{#1}\@@href}%
\providecommand \@@href[1]{\endgroup#1\@@endlink}%
\providecommand \@sanitize@url [0]{\catcode `\\12\catcode `\$12\catcode
  `\&12\catcode `\#12\catcode `\^12\catcode `\_12\catcode `\%12\relax}%
\providecommand \@@startlink[1]{}%
\providecommand \@@endlink[0]{}%
\providecommand \url  [0]{\begingroup\@sanitize@url \@url }%
\providecommand \@url [1]{\endgroup\@href {#1}{\urlprefix }}%
\providecommand \urlprefix  [0]{URL }%
\providecommand \Eprint [0]{\href }%
\providecommand \doibase [0]{http://dx.doi.org/}%
\providecommand \selectlanguage [0]{\@gobble}%
\providecommand \bibinfo  [0]{\@secondoftwo}%
\providecommand \bibfield  [0]{\@secondoftwo}%
\providecommand \translation [1]{[#1]}%
\providecommand \BibitemOpen [0]{}%
\providecommand \bibitemStop [0]{}%
\providecommand \bibitemNoStop [0]{.\EOS\space}%
\providecommand \EOS [0]{\spacefactor3000\relax}%
\providecommand \BibitemShut  [1]{\csname bibitem#1\endcsname}%
\let\auto@bib@innerbib\@empty
\end{thebibliography}%


\begin{thebibliography}{132}%
\makeatletter
\providecommand \@ifxundefined [1]{%
 \@ifx{#1\undefined}
}%
\providecommand \@ifnum [1]{%
 \ifnum #1\expandafter \@firstoftwo
 \else \expandafter \@secondoftwo
 \fi
}%
\providecommand \@ifx [1]{%
 \ifx #1\expandafter \@firstoftwo
 \else \expandafter \@secondoftwo
 \fi
}%
\providecommand \natexlab [1]{#1}%
\providecommand \enquote  [1]{``#1''}%
\providecommand \bibnamefont  [1]{#1}%
\providecommand \bibfnamefont [1]{#1}%
\providecommand \citenamefont [1]{#1}%
\providecommand \href@noop [0]{\@secondoftwo}%
\providecommand \href [0]{\begingroup \@sanitize@url \@href}%
\providecommand \@href[1]{\@@startlink{#1}\@@href}%
\providecommand \@@href[1]{\endgroup#1\@@endlink}%
\providecommand \@sanitize@url [0]{\catcode `\\12\catcode `\$12\catcode
  `\&12\catcode `\#12\catcode `\^12\catcode `\_12\catcode `\%12\relax}%
\providecommand \@@startlink[1]{}%
\providecommand \@@endlink[0]{}%
\providecommand \url  [0]{\begingroup\@sanitize@url \@url }%
\providecommand \@url [1]{\endgroup\@href {#1}{\urlprefix }}%
\providecommand \urlprefix  [0]{URL }%
\providecommand \Eprint [0]{\href }%
\providecommand \doibase [0]{http://dx.doi.org/}%
\providecommand \selectlanguage [0]{\@gobble}%
\providecommand \bibinfo  [0]{\@secondoftwo}%
\providecommand \bibfield  [0]{\@secondoftwo}%
\providecommand \translation [1]{[#1]}%
\providecommand \BibitemOpen [0]{}%
\providecommand \bibitemStop [0]{}%
\providecommand \bibitemNoStop [0]{.\EOS\space}%
\providecommand \EOS [0]{\spacefactor3000\relax}%
\providecommand \BibitemShut  [1]{\csname bibitem#1\endcsname}%
\let\auto@bib@innerbib\@empty

\bibitem [{\citenamefont {Bednorz}\ \emph {et~al.}(1986)\citenamefont {Bednorz},
  \ and\ \citenamefont{Müller}}]{bednorz_1986}%
  \BibitemOpen
  \bibfield  {author} {\bibinfo {author} {\bibfnamefont {J.~G.}~\bibnamefont
  {Bednorz}}, \and\ \bibinfo {author} {\bibfnamefont {K.~A.}\ \bibnamefont {Müller}},\
  }\bibinfo {title} {Possible high $T_c$ superconductivity in the Ba{-}La{-}Cu{-}O system}, \href {\doibase 10.1007/BF01303701} {\bibfield  {journal} {\bibinfo
  {journal} {Zeitschrift für Physik B Condensed Matter}\ }\textbf {\bibinfo {volume} {64}},\ \bibinfo
  {pages} {189} (\bibinfo {year} {1986})}\BibitemShut {NoStop}%
\bibitem [{\citenamefont {Anderson}(1987)}]{anderson_1987}%
  \BibitemOpen
  \bibfield  {author} {\bibinfo {author} {\bibfnamefont {P.~W.}~\bibnamefont
  {Anderson}},\ }\bibinfo {title} {The Resonating Valence Bond State in $\text{La}_2 \text{Cu} \text{O}_4$ and Superconductivity}, \href {\doibase 10.1126/science.235.4793.1196} {\bibfield  {journal} {\bibinfo
   {journal} {Science}\ }\textbf {\bibinfo {volume} {235}},\ \bibinfo
  {number} {4793} (\bibinfo {year} {1987})}\BibitemShut {NoStop}%
\bibitem [{\citenamefont {Zhang}\ and\ \citenamefont
  {Rice}(1988)}]{zhang_1988}%
  \BibitemOpen
  \bibfield  {author} {\bibinfo {author} {\bibfnamefont {F.~C.}~\bibnamefont
  {Zhang}}\ and\ \bibinfo {author} {\bibfnamefont {T.~M.}~\bibnamefont {Rice}},\
  }\bibinfo {title} {Effective Hamiltonian for the superconducting Cu oxides}, \href {\doibase 10.1103/PhysRevB.37.3759} {\bibfield  {journal} {\bibinfo
  {journal} {Phys. Rev. B}\ }\textbf {\bibinfo {volume} {37}},\ \bibinfo
  {pages} {3759} (\bibinfo {year} {1988})}\BibitemShut {NoStop}%
\bibitem [{\citenamefont {Ogata}\ \emph {et~al.}(1991)\citenamefont {Ogata},
  \citenamefont {Luchini}, \citenamefont {Sorella},\ and\ \citenamefont
  {Assaad}}]{ogata_phase_1991}%
  \BibitemOpen
  \bibfield  {author} {\bibinfo {author} {\bibfnamefont {M.}~\bibnamefont
  {Ogata}}, \bibinfo {author} {\bibfnamefont {M.~U.}\ \bibnamefont {Luchini}},
  \bibinfo {author} {\bibfnamefont {S.}~\bibnamefont {Sorella}}, \ and\
  \bibinfo {author} {\bibfnamefont {F.~F.}\ \bibnamefont {Assaad}},\ }\bibinfo {title} {Phase Diagram of the Two-Dimensional $t$-$J$ Model}, \href
  {\doibase 10.1103/PhysRevLett.66.2388} {\bibfield  {journal} {\bibinfo
  {journal} {Phys. Rev. Lett.}\ }\textbf {\bibinfo {volume} {66}},\
  \bibinfo {pages} {2388} (\bibinfo {year} {1991})}\BibitemShut {NoStop}%
\bibitem [{\citenamefont {Moreno}\ \emph {et~al.}(2011)\citenamefont {Moreno},
  \citenamefont {Muramatsu},\ and\ \citenamefont
  {Manmana}}]{moreno_ground-state_2011}%
  \BibitemOpen
  \bibfield  {author} {\bibinfo {author} {\bibfnamefont {A.}~\bibnamefont
  {Moreno}}, \bibinfo {author} {\bibfnamefont {A.}~\bibnamefont {Muramatsu}}, \
  and\ \bibinfo {author} {\bibfnamefont {S.~R.}\ \bibnamefont {Manmana}},\
  }\bibinfo {title} {Ground-state phase diagram of the two-dimensional $t$-$J$ model}, \href {\doibase 10.1103/PhysRevB.83.205113} {\bibfield  {journal} {\bibinfo
  {journal} {Phys. Rev. B}\ }\textbf {\bibinfo {volume} {83}},\ \bibinfo
  {pages} {205113} (\bibinfo {year} {2011})}\BibitemShut {NoStop}%
\bibitem [{\citenamefont {Gross}\ and\ \citenamefont{Wilczek}(2018)}]{gross_1973}%
  \BibitemOpen
  \bibfield  {author} {\bibinfo {author} {\bibfnamefont {D.~J.}~\bibnamefont
  {Gross}}\ and\ \bibinfo {author} {\bibfnamefont {F.}~\bibnamefont {Wilczek}},\
  }\bibinfo {title} {Ultraviolet Behavior of Non-Abelian Gauge Theories}, \href {\doibase 10.1103/PhysRevLett.30.1343} {\bibfield  {journal} {\bibinfo
  {journal} {Phys. Rev. Lett.}\ }\textbf {\bibinfo {volume} {30}},\ \bibinfo
  {pages} {1343} (\bibinfo {year} {1973})}\BibitemShut {NoStop}%
\bibitem [{\citenamefont {Politzer}(1973)}]{politzer_1973}%
  \BibitemOpen
  \bibfield  {author} {\bibinfo {author} {\bibfnamefont {H.~D.}~\bibnamefont
  {Politzer}},\ }\bibinfo {title} {Reliable Perturbative Results for Strong Interactions?}, \href {\doibase 10.1103/PhysRevLett.30.1346} {\bibfield  {journal} {\bibinfo
   {journal} {Phys. Rev. Lett.}\ }\textbf {\bibinfo {volume} {30}},\ \bibinfo
  {pages} {1346} (\bibinfo {year} {1973})}\BibitemShut {NoStop}%
\bibitem [{\citenamefont {Yuan}\ and\ \citenamefont
  {Fu}(2018)}]{yuan_2018}%
  \BibitemOpen
  \bibfield  {author} {\bibinfo {author} {\bibfnamefont {N.~F.~Q.}~\bibnamefont
  {Yuan}}\ and\ \bibinfo {author} {\bibfnamefont {L.}~\bibnamefont {Fu}},\
  }\bibinfo {title} {Model for the metal-insulator transition in graphene superlattices and beyond}, \href {\doibase 10.1103/PhysRevB.98.045103} {\bibfield  {journal} {\bibinfo
  {journal} {Phys. Rev. B}\ }\textbf {\bibinfo {volume} {98}},\ \bibinfo
  {pages} {045103} (\bibinfo {year} {2018})}\BibitemShut {NoStop}%
\bibitem [{\citenamefont {Bloch}\ \emph {et~al.}(2012)\citenamefont {Bloch},
 \citenamefont {Dalibard},\ and\ \citenamefont
 {Nascimbène}}]{bloch_quantum_2012}%
 \BibitemOpen
 \bibfield  {author} {\bibinfo {author} {\bibfnamefont {I.}~\bibnamefont
 {Bloch}}, \bibinfo {author} {\bibfnamefont {J.}~\bibnamefont {Dalibard}}, \
 and\ \bibinfo {author} {\bibfnamefont {S.}~\bibnamefont {Nascimbène}},\
 }\bibinfo {title} {Quantum simulations with ultracold quantum gases}, \href {\doibase 10.1038/nphys2259} {\bibfield  {journal} {\bibinfo
 {journal} {Nature Physics}\ }\textbf {\bibinfo {volume} {8}},\ \bibinfo
 {pages} {267} (\bibinfo {year} {2012})}\BibitemShut {NoStop}%
\bibitem [{\citenamefont {Bloch}(2005)}]{bloch_ultracold_2005}%
  \BibitemOpen
  \bibfield  {author} {\bibinfo {author} {\bibfnamefont {I.}~\bibnamefont
  {Bloch}},\ }\bibinfo {title} {Ultracold quantum gases in optical lattices}, \href {\doibase 10.1038/nphys138} {\bibfield  {journal} {\bibinfo
   {journal} {Nature Physics}\ }\textbf {\bibinfo {volume} {1}},\ \bibinfo
  {pages} {23} (\bibinfo {year} {2005})}\BibitemShut {NoStop}%
\bibitem [{\citenamefont {Zhang}\ \emph {et~al.}(2020)\citenamefont {Zhang},
  \citenamefont {Cheng}, \citenamefont {Zhang},\ and\ \citenamefont
  {Zhai}}]{zhang_controlling_2020}%
  \BibitemOpen
  \bibfield  {author} {\bibinfo {author} {\bibfnamefont {R.}~\bibnamefont
  {Zhang}}, \bibinfo {author} {\bibfnamefont {Y.}~\bibnamefont {Cheng}},
  \bibinfo {author} {\bibfnamefont {P.}~\bibnamefont {Zhang}}, \ and\ \bibinfo
  {author} {\bibfnamefont {H.}~\bibnamefont {Zhai}},\ }\bibinfo {title} {Controlling the interaction of ultracold alkaline-earth atoms}, \href {\doibase 10.1038/s42254-020-0157-9} {\bibfield  {journal} {\bibinfo  {journal} {Nature
  Reviews Physics}\ }\textbf {\bibinfo {volume} {2}},\ \bibinfo {pages} {213}
  (\bibinfo {year} {2020})}\BibitemShut {NoStop}%
\bibitem [{\citenamefont {Blatt}\ and\ \citenamefont
  {Roos}(2012)}]{blatt_quantum_2012}%
  \BibitemOpen
  \bibfield  {author} {\bibinfo {author} {\bibfnamefont {R.}~\bibnamefont
  {Blatt}}\ and\ \bibinfo {author} {\bibfnamefont {C.~F.}\ \bibnamefont
  {Roos}},\ }\bibinfo {title} {Quantum simulations with trapped ions}, \href {\doibase 10.1038/nphys2252} {\bibfield  {journal} {\bibinfo
   {journal} {Nature Physics}\ }\textbf {\bibinfo {volume} {8}},\ \bibinfo
  {pages} {277} (\bibinfo {year} {2012})}\BibitemShut {NoStop}%
\bibitem [{\citenamefont {Tusi}\ \emph {et~al.}(2021)\citenamefont {Tusi},
  \citenamefont {Franchi}, \citenamefont {Livi}, \citenamefont {Baumann},
  \citenamefont {Orenes}, \citenamefont {Del~Re}, \citenamefont {Barfknecht},
  \citenamefont {Zhou}, \citenamefont {Inguscio}, \citenamefont {Cappellini},
  \citenamefont {Capone}, \citenamefont {Catani},\ and\ \citenamefont
  {Fallani}}]{tusi_flavour-selective_2021}%
  \BibitemOpen
  \bibfield  {author} {\bibinfo {author} {\bibfnamefont {D.}~\bibnamefont
  {Tusi}}, \bibinfo {author} {\bibfnamefont {L.}~\bibnamefont {Franchi}},
  \bibinfo {author} {\bibfnamefont {L.~F.}\ \bibnamefont {Livi}}, \bibinfo
  {author} {\bibfnamefont {K.}~\bibnamefont {Baumann}}, \bibinfo {author}
  {\bibfnamefont {D.~B.}\ \bibnamefont {Orenes}}, \bibinfo {author}
  {\bibfnamefont {L.}~\bibnamefont {Del~Re}}, \bibinfo {author} {\bibfnamefont
  {R.~E.}\ \bibnamefont {Barfknecht}}, \bibinfo {author} {\bibfnamefont
  {T.}~\bibnamefont {Zhou}}, \bibinfo {author} {\bibfnamefont {M.}~\bibnamefont
  {Inguscio}}, \bibinfo {author} {\bibfnamefont {G.}~\bibnamefont
  {Cappellini}}, \bibinfo {author} {\bibfnamefont {M.}~\bibnamefont {Capone}},
  \bibinfo {author} {\bibfnamefont {J.}~\bibnamefont {Catani}}, \ and\ \bibinfo
  {author} {\bibfnamefont {L.}~\bibnamefont {Fallani}},\ }\bibinfo {title} {Flavour-selective localization in interacting lattice fermions}, \href
  {\doibase 10.1038/s41567-022-01726-5} {\bibfield  {journal} {\bibinfo  {journal}
  {Nature Physics}\ } \textbf {\bibinfo {volume} {18}},\ \bibinfo {pages} {1201} (\bibinfo
  {year} {2022})}\BibitemShut {NoStop}%
\bibitem [{\citenamefont {Taie}\ \emph {et~al.}(2012)\citenamefont {Taie},
  \citenamefont {Yamazaki}, \citenamefont {Sugawa},\ and\ \citenamefont
  {Takahashi}}]{taie_su6_2012}%
  \BibitemOpen
  \bibfield  {author} {\bibinfo {author} {\bibfnamefont {S.}~\bibnamefont {Taie}}, \bibinfo {author} {\bibfnamefont {R.}~\bibnamefont {Yamazaki}}, \bibinfo {author} {\bibfnamefont {S.}~\bibnamefont {Sugawa}}, \ and\ \bibinfo{author} {\bibfnamefont {Y.}~\bibnamefont {Takahashi}},\ }\bibinfo {title} {An $\rm{SU}\left(6\right)$ Mott insulator of an atomic Fermi gas realized by large-spin Pomeranchuk cooling}, \href {\doibase 10.1038/nphys2430} {\bibfield  {journal} {\bibinfo  {journal} {Nature Physics}\ }\textbf {\bibinfo {volume} {8}},\ \bibinfo {pages} {825} (\bibinfo{year} {2012})}\BibitemShut {NoStop}%
\bibitem [{\citenamefont {Taie}\ \emph {et~al.}(2022)\citenamefont {Taie},
 \citenamefont {Ibarra-García-Padillaa}, \citenamefont {Nishizawa},
 \citenamefont {Takasu}, \citenamefont {Kuno}, \citenamefont {Wei},
 \citenamefont {Scalettar}, \citenamefont {Hazzard},\ and\ \citenamefont
 {Takahashi}}]{taie_observation_2022}%
 \BibitemOpen
 \bibfield  {author} {\bibinfo {author} {\bibfnamefont {S.}~\bibnamefont
 {Taie}}, \bibinfo {author} {\bibfnamefont {E.}~\bibnamefont
 {Ibarra-García-Padillaa}}, \bibinfo {author} {\bibfnamefont {N.}~\bibnamefont
 {Nishizawa}}, \bibinfo {author} {\bibfnamefont {Y.}~\bibnamefont {Takasu}},
 \bibinfo {author} {\bibfnamefont {Y.}~\bibnamefont {Kuno}}, \bibinfo {author}
 {\bibfnamefont {H.~T.}\ \bibnamefont {Wei}}, \bibinfo {author} {\bibfnamefont
 {R.~T.}\ \bibnamefont {Scalettar}}, \bibinfo {author} {\bibfnamefont {K.~R.}\
 \bibnamefont {Hazzard}}, \ and\ \bibinfo {author} {\bibfnamefont
 {Y.}~\bibnamefont {Takahashi}},\ }\bibinfo {title} {Observation of antiferromagnetic correlations in a 1D $\rm{SU}\left(N\right)$ Hubbard model}, \href
 {\doibase 10.1038/s41567-022-01725-6} {\bibfield  {journal} {\bibinfo  {journal}
 {Nature Physics}\ } \textbf {\bibinfo {volume} {18}},\ \bibinfo {pages} {1356} (\bibinfo
 {year} {2022})}\BibitemShut {NoStop}%
\bibitem [{\citenamefont {Capponi}\ \emph {et~al.}(2016)\citenamefont
  {Capponi}, \citenamefont {Lecheminant},\ and\ \citenamefont
  {Totsuka}}]{capponi_phases_2016}%
  \BibitemOpen
  \bibfield  {author} {\bibinfo {author} {\bibfnamefont {S.}~\bibnamefont
  {Capponi}}, \bibinfo {author} {\bibfnamefont {P.}~\bibnamefont
  {Lecheminant}}, \ and\ \bibinfo {author} {\bibfnamefont {K.}~\bibnamefont
  {Totsuka}},\ }\bibinfo {title} {Phases of one-dimensional $\rm{SU}\left(N\right)$ cold atomic Fermi gases—From molecular Luttinger liquids to topological phases}, \href {\doibase 10.1016/j.aop.2016.01.011} {\bibfield
  {journal} {\bibinfo  {journal} {Annals of Physics}\ }\textbf {\bibinfo
  {volume} {367}},\ \bibinfo {pages} {50} (\bibinfo {year} {2016})}\BibitemShut {NoStop}%
\bibitem [{\citenamefont {Giamarchi}(2004)}]{giamarchi_quantum_2004}%
  \BibitemOpen
  \bibfield  {author} {\bibinfo {author} {\bibfnamefont {T.}~\bibnamefont
  {Giamarchi}},\ }\href@noop {} {\emph {\bibinfo {title} {Quantum physics in
  one dimension}}},\ \bibinfo {series} {The international series of monographs
  on physics}\ No.\ \bibinfo {number} {121}\ (\bibinfo  {publisher} {Clarendon
  ; Oxford University Press},\ \bibinfo {address} {Oxford : New York},\
  \bibinfo {year} {2004})\BibitemShut
  {NoStop}%
\bibitem [{\citenamefont {Gogolin}\ \emph {et~al.}(1998)\citenamefont
  {Gogolin}, \citenamefont {Nersesyan},\ and\ \citenamefont
  {Tsvelik}}]{gogolin_bosonization_1998}%
  \BibitemOpen
  \bibfield  {author} {\bibinfo {author} {\bibfnamefont {A.~O.}\ \bibnamefont
  {Gogolin}}, \bibinfo {author} {\bibfnamefont {A.~A.}\ \bibnamefont
  {Nersesyan}}, \ and\ \bibinfo {author} {\bibfnamefont {A.~M.}\ \bibnamefont
  {Tsvelik}},\ }\href@noop {} {\emph {\bibinfo {title} {Bosonization and
  strongly correlated systems}}}\ (\bibinfo  {publisher} {Cambridge University
  Press},\ \ \bibinfo
  {year} {1998})\BibitemShut {NoStop}%
\bibitem [{\citenamefont {Schlottmann}(1993)}]{schlottmann_groundstate_1993}%
  \BibitemOpen
  \bibfield  {author} {\bibinfo {author} {\bibfnamefont {P.}~\bibnamefont
  {Schlottmann}},\ }\bibinfo {title} {Groundstate properties of the degenerate supersymmetric $t$-$J$ model with crystalline field splitting in one dimension }, \href {\doibase 10.1063/1.352541} {\bibfield  {journal}
  {\bibinfo  {journal} {J. App. Phys.}\ }\textbf {\bibinfo
  {volume} {73}},\ \bibinfo {pages} {6645} (\bibinfo {year}
  {1993})}\BibitemShut {NoStop}%
\bibitem [{\citenamefont {Itoi}\ \emph {et~al.}(2000)\citenamefont {Itoi},
  \citenamefont {Qin},\ and\ \citenamefont {Affleck}}]{itoi_phase_2000}%
  \BibitemOpen
  \bibfield  {author} {\bibinfo {author} {\bibfnamefont {C.}~\bibnamefont
  {Itoi}}, \bibinfo {author} {\bibfnamefont {S.}~\bibnamefont {Qin}}, \ and\
  \bibinfo {author} {\bibfnamefont {I.}~\bibnamefont {Affleck}},\ }\bibinfo {title} {Phase transitions in anisotropic $\rm{SU}\left(N\right)$ spin chains}, \href
  {\doibase 10.1103/PhysRevB.61.6747} {\bibfield  {journal} {\bibinfo
  {journal} {Phys. Rev. B}\ }\textbf {\bibinfo {volume} {61}},\ \bibinfo
  {pages} {6747} (\bibinfo {year} {2000})}\BibitemShut {NoStop}%
\bibitem [{\citenamefont {Affleck}(1986)}]{affleck_exact_1986}%
  \BibitemOpen
  \bibfield  {author} {\bibinfo {author} {\bibfnamefont {I.}~\bibnamefont
  {Affleck}},\ }\bibinfo {title} {Exact critical exponents for quantum spin chains}, \href {\doibase 10.1016/0550-3213(86)90167-7} {\bibfield
  {journal} {\bibinfo  {journal} {Nuclear Physics B}\ }\textbf {\bibinfo
  {volume} {265}},\ \bibinfo {pages} {409} (\bibinfo {year}
  {1986})},  \bibinfo {title} {Critical behaviour of $\rm{SU}\left(n\right)$ quantum chains and topological non-linear $\sigma$-models}, \href {\doibase 10.1016/0550-3213(88)90117-4} {\bibfield
  {journal} {\bibinfo  {journal} {Nuclear Physics B}\ }\textbf {\bibinfo
  {volume} {305}},\ \bibinfo {pages} {582} (\bibinfo {year}
  {1988})}\BibitemShut {NoStop}%
\bibitem [{\citenamefont {Knizhnik}\ and\ \citenamefont
  {Zamolodchikov}(1984)}]{knizhnik_current_1984}%
  \BibitemOpen
  \bibfield  {author} {\bibinfo {author} {\bibfnamefont {V.}~\bibnamefont
  {Knizhnik}}\ and\ \bibinfo {author} {\bibfnamefont {A.}~\bibnamefont
  {Zamolodchikov}},\ }\bibinfo {title} {Current algebra and Wess-Zumino model in two dimensions}, \href {\doibase 10.1016/0550-3213(84)90374-2} {\bibfield
  {journal} {\bibinfo  {journal} {Nuclear Physics B}\ }\textbf {\bibinfo
  {volume} {247}},\ \bibinfo {pages} {83} (\bibinfo {year} {1984})}\BibitemShut
  {NoStop}%
\bibitem [{\citenamefont {Di~Francesco}\ \emph {et~al.}(1997)\citenamefont
 {Di~Francesco}, \citenamefont {Mathieu},\ and\ \citenamefont
 {Sénéchal}}]{di_francesco_conformal_1997}%
 \BibitemOpen
 \bibfield  {author} {\bibinfo {author} {\bibfnamefont {P.}~\bibnamefont
 {Di~Francesco}}, \bibinfo {author} {\bibfnamefont {P.}~\bibnamefont
 {Mathieu}}, \ and\ \bibinfo {author} {\bibfnamefont {D.}~\bibnamefont
 {Sénéchal}},\ }\href@noop {} {\emph {\bibinfo {title} {Conformal field
 theory}}},\ (\bibinfo  {publisher}
 {Springer},\ \bibinfo {address} {New York},\ \bibinfo {year}
 {1997})\BibitemShut {NoStop}%
\bibitem [{\citenamefont {Sutherland}(1975)}]{sutherland_model_1975}%
  \BibitemOpen
  \bibfield  {author} {\bibinfo {author} {\bibfnamefont {B.}~\bibnamefont
  {Sutherland}},\ }\bibinfo {title} {Model for the Mott transition in the two-dimensional Hubbard model}, \href {\doibase 10.1103/PhysRevB.12.3795} {\bibfield
  {journal} {\bibinfo  {journal} {Phys. Rev. B}\ }\textbf {\bibinfo
  {volume} {12}},\ \bibinfo {pages} {3795} (\bibinfo {year}
  {1975})}\BibitemShut {NoStop}%
\bibitem [{\citenamefont {Nataf}\ and\ \citenamefont
  {Mila}(2016)}]{nataf_exact_2016}%
  \BibitemOpen
  \bibfield  {author} {\bibinfo {author} {\bibfnamefont {P.}~\bibnamefont
  {Nataf}}\ and\ \bibinfo {author} {\bibfnamefont {F.}~\bibnamefont {Mila}},\
  }\bibinfo {title} {Exact diagonalization of the t-J model on small clusters}, \href {\doibase10.1103/PhysRevB.93.155134} {\bibfield  {journal} {\bibinfo
  {journal} {Phys. Rev. B}\ }\textbf {\bibinfo {volume} {93}},\ \bibinfo
  {pages} {155134} (\bibinfo {year} {2016})}\BibitemShut {NoStop}%
\bibitem [{\citenamefont {Wan}\ \emph {et~al.}(2017)\citenamefont {Wan},
  \citenamefont {Nataf},\ and\ \citenamefont {Mila}}]{wan_exact_2017}%
  \BibitemOpen
  \bibfield  {author} {\bibinfo {author} {\bibfnamefont {K.}~\bibnamefont
  {Wan}}, \bibinfo {author} {\bibfnamefont {P.}~\bibnamefont {Nataf}}, \ and\
  \bibinfo {author} {\bibfnamefont {F.}~\bibnamefont {Mila}},\ }\bibinfo {title} {Exact diagonalization of $\rm{SU}\left(N\right)$ Heisenberg and Affleck-Kennedy-Lieb-Tasaki chains using the full $\rm{SU}\left(N\right)$ symmetry}, \href {\doibase 10.1103/PhysRevB.96.115159} {\bibfield  {journal} {\bibinfo  {journal}
  {Phys. Rev. B}\ }\textbf {\bibinfo {volume} {96}},\ \bibinfo {pages} {115159}
  (\bibinfo {year} {2017})}\BibitemShut {NoStop}%
\bibitem [{\citenamefont {Vörös}\ and\ \citenamefont
  {Penc}(2021)}]{vörös_2021}%
  \BibitemOpen
  \bibfield  {author} {\bibinfo {author} {\bibfnamefont {D.}~\bibnamefont
  {Vörös}}\ and\ \bibinfo {author} {\bibfnamefont {K.}~\bibnamefont {Penc}},\
  }\bibinfo {title} {Magnetic properties of the t-J model on a square lattice}, \href {\doibase 10.1103/PhysRevB.104.184426} {\bibfield  {journal} {\bibinfo
  {journal} {Phys. Rev. B}\ }\textbf {\bibinfo {volume} {104}},\ \bibinfo
  {pages} {184426} (\bibinfo {year} {2021})}\BibitemShut {NoStop}%
\bibitem [{\citenamefont {Nataf}\ and\ \citenamefont
  {Mila}(2018)}]{nataf_density_2018}%
  \BibitemOpen
  \bibfield  {author} {\bibinfo {author} {\bibfnamefont {P.}~\bibnamefont
  {Nataf}}\ and\ \bibinfo {author} {\bibfnamefont {F.}~\bibnamefont {Mila}},\
  }\bibinfo {title} {Density matrix renormalization group simulations of $\rm{SU}\left(N\right)$ Heisenberg chains using standard Young tableaus: Fundamental representation and comparison with a finite-size Bethe ansatz}, \href {\doibase 10.1103/PhysRevB.97.134420} {\bibfield  {journal} {\bibinfo
  {journal} {Phys. Rev. B}\ }\textbf {\bibinfo {volume} {97}},\ \bibinfo
  {pages} {134420} (\bibinfo {year} {2018})}\BibitemShut {NoStop}%
\bibitem [{\citenamefont {Führinger}\ \emph {et~al.}(2008)\citenamefont
 {Führinger}, \citenamefont {Rachel}, \citenamefont {Thomale}, \citenamefont
 {Greiter},\ and\ \citenamefont {Schmitteckert}}]{fuhringer_dmrg_2008}%
 \BibitemOpen
 \bibfield  {author} {\bibinfo {author} {\bibfnamefont {M.}~\bibnamefont
 {Führinger}}, \bibinfo {author} {\bibfnamefont {S.}~\bibnamefont {Rachel}},
 \bibinfo {author} {\bibfnamefont {R.}~\bibnamefont {Thomale}}, \bibinfo
 {author} {\bibfnamefont {M.}~\bibnamefont {Greiter}}, \ and\ \bibinfo
 {author} {\bibfnamefont {P.}~\bibnamefont {Schmitteckert}},\ }\bibinfo {title} {DMRG studies of critical $\rm{SU}\left(N\right)$ spin chains}, \href {\doibase 10.1002/andp.20085201204} {\bibfield  {journal} {\bibinfo  {journal} {Annalen der Physik
 }\ }\textbf {\bibinfo {volume} {12}},\ \bibinfo {pages}
 {922-936} (\bibinfo {year} {2008})}\BibitemShut {NoStop}%
\bibitem [{\citenamefont {Fromholz}\ \emph {et~al.}(2019)\citenamefont {Fromholz},
 \citenamefont {Capponi}, \citenamefont {Lecheminant}, \citenamefont {Papoular},\
 and\ \citenamefont {Totsuka}}]{fromholz_2019}%
 \BibitemOpen
 \bibfield  {author} {\bibinfo {author} {\bibfnamefont {P.}~\bibnamefont
 {Fromholz}}, \bibinfo {author} {\bibfnamefont {S.}~\bibnamefont {Capponi}},
 \bibinfo {author} {\bibfnamefont {P.}~\bibnamefont {Lecheminant}},
 \bibinfo {author} {\bibfnamefont {D. J.}~\bibnamefont {Papoular}},\ and\
 \bibinfo {author} {\bibfnamefont {K.}~\bibnamefont {Totsuka}},\ }\bibinfo {title} {Haldane phases with ultracold fermionic atoms in double-well optical lattices}, \href {\doibase 10.1103/PhysRevB.99.054414} {\bibfield  {journal} {\bibinfo  {journal} {Phys. Rev. B}\ }\textbf {\bibinfo {volume} {99}},\ \bibinfo {pages} {054414} (\bibinfo {year} {2019})}\BibitemShut {NoStop}%
\bibitem [{\citenamefont {Gozel}\ \emph {et~al.}(2021)\citenamefont {Gozel}, \citenamefont {Nataf},\ and\ \citenamefont {Mila}}]{gozel_2020}%
 \BibitemOpen
 \bibfield  {author} {\bibinfo {author} {\bibfnamefont {S.}~\bibnamefont {Gozel}}, \bibinfo {author} {\bibfnamefont {P.}~\bibnamefont {Nataf}}, \ and\ \bibinfo {author} {\bibfnamefont {F.}~\bibnamefont {Mila}},\ }\bibinfo {title} {Haldane Gap of the Three-Box Symmetric $\rm{SU}\left(3\right)$ Chain}, \href {\doibase 10.1103/PhysRevLett.125.057202} {\bibfield  {journal} {\bibinfo  {journal} {Phys. Rev. Lett.}\ }\textbf {\bibinfo {volume} {125}},\ \bibinfo {pages} {057202} (\bibinfo {year} {2020})}\BibitemShut {NoStop}%
\bibitem [{\citenamefont {Nataf}\ \emph {et~al.}(2021)\citenamefont {Nataf},
 \citenamefont {Gozel},\ and\ \citenamefont {Mila}}]{nataf_edge_2021}%
 \BibitemOpen
 \bibfield  {author} {\bibinfo {author} {\bibfnamefont {P.}~\bibnamefont
 {Nataf}}, \bibinfo {author} {\bibfnamefont {S.}~\bibnamefont {Gozel}}, \ and\
 \bibinfo {author} {\bibfnamefont {F.}~\bibnamefont {Mila}},\ }\ \bibinfo {title} {Edge states and universality class of the critical two-box symmetric $\rm{SU}\left(3\right)$ chain}, \href {\doibase 10.1103/PhysRevB.104.L180411} {\bibfield  {journal} {\bibinfo  {journal} {Phys. Rev. B}\ }\textbf {\bibinfo {volume} {104}},\ \bibinfo {pages} {L180411}
 (\bibinfo {year} {2021})}\BibitemShut {NoStop}%
\bibitem [{\citenamefont {Assaraf}\ \emph {et~al.}(1999)\citenamefont
  {Assaraf}, \citenamefont {Azaria}, \citenamefont {Caffarel},\ and\
  \citenamefont {Lecheminant}}]{assaraf_metal-insulator_1999}%
  \BibitemOpen
  \bibfield  {author} {\bibinfo {author} {\bibfnamefont {R.}~\bibnamefont
  {Assaraf}}, \bibinfo {author} {\bibfnamefont {P.}~\bibnamefont {Azaria}},
  \bibinfo {author} {\bibfnamefont {M.}~\bibnamefont {Caffarel}}, \ and\
  \bibinfo {author} {\bibfnamefont {P.}~\bibnamefont {Lecheminant}},\ }\bibinfo {title} {Metal-insulator transition in the one-dimensional $\rm{SU}\left(N\right)$ Hubbard model}, \href
  {\doibase 10.1103/PhysRevB.60.2299} {\bibfield  {journal} {\bibinfo
  {journal} {Phys. Rev. B}\ }\textbf {\bibinfo {volume} {60}},\ \bibinfo
  {pages} {2299} (\bibinfo {year} {1999})}\BibitemShut {NoStop}%
\bibitem [{\citenamefont {he}(2022)}]{he_2022}%
  \BibitemOpen
  \bibfield  {author} {\bibinfo {author} {\bibfnamefont {J. C.}~\bibnamefont
  {He}}, \bibinfo {author} {\bibfnamefont {J.}\ \bibnamefont
  {Hou}}\ and\ \bibinfo {author} {\bibfnamefont {Y.}\ \bibnamefont
  {Chen}},\ }\bibinfo {title} {Gutzwiller approximation approach to the $\rm{SU}\left(4\right)$ $t$-$J$ model}, \href {\doibase 10.1103/PhysRevB.105.245117} {\bibfield
  {journal} {\bibinfo  {journal} {Phys. Rev. B}\ }\textbf {\bibinfo {volume} {105}},\ \bibinfo {pages} {245117}
  (\bibinfo {year} {2022})}\BibitemShut {NoStop}%
\bibitem [{\citenamefont {Fishman}\ \emph {et~al.}(2020)\citenamefont
  {Fishman}, \citenamefont {White},\ and\ \citenamefont
  {Stoudenmire}}]{fishman_itensor_2020}%
  \BibitemOpen
  \bibfield  {author} {\bibinfo {author} {\bibfnamefont {M.}~\bibnamefont
  {Fishman}}, \bibinfo {author} {\bibfnamefont {S.~R.}\ \bibnamefont {White}},
  \ and\ \bibinfo {author} {\bibfnamefont {E.~M.}\ \bibnamefont
  {Stoudenmire}},\ }\bibinfo {title} {The ITensor Software Library for Tensor Network Calculations}, \href {\doibase 10.21468/SciPostPhysCodeb.4} {\bibfield
  {journal} {\bibinfo  {journal} {SciPost Phys. Codebases}\ }\bibinfo
  {pages} {4} (\bibinfo {year} {2022})}
  \BibitemShut {NoStop}%
\bibitem [{\citenamefont {Clay}\ \emph {et~al.}(1999)\citenamefont {Clay},
  \citenamefont {Sandvik},\ and\ \citenamefont
  {Campbell}}]{clay_possible_1999}%
  \BibitemOpen
  \bibfield  {author} {\bibinfo {author} {\bibfnamefont {R.~T.}\ \bibnamefont
  {Clay}}, \bibinfo {author} {\bibfnamefont {A.~W.}\ \bibnamefont {Sandvik}}, \
  and\ \bibinfo {author} {\bibfnamefont {D.~K.}\ \bibnamefont {Campbell}},\
  }\bibinfo {title} {Possible exotic phases in the one-dimensional extended Hubbard model}, \href {\doibase 10.1103/PhysRevB.59.4665} {\bibfield  {journal} {\bibinfo
  {journal} {Phys. Rev. B}\ }\textbf {\bibinfo {volume} {59}},\ \bibinfo
  {pages} {4665} (\bibinfo {year} {1999})}\BibitemShut {NoStop}%
\bibitem [{\citenamefont {Ejima}\ \emph {et~al.}(2005)\citenamefont {Ejima},
  \citenamefont {Gebhard},\ and\ \citenamefont
  {Nishimoto}}]{ejima_tomonaga-luttinger_2005}%
  \BibitemOpen
  \bibfield  {author} {\bibinfo {author} {\bibfnamefont {S.}~\bibnamefont
  {Ejima}}, \bibinfo {author} {\bibfnamefont {F.}~\bibnamefont {Gebhard}}, \
  and\ \bibinfo {author} {\bibfnamefont {S.}~\bibnamefont {Nishimoto}},\ }\bibinfo {title} {Tomonaga-Luttinger parameters for doped Mott insulators}, \href
  {\doibase 10.1209/epl/i2005-10020-8} {\bibfield  {journal} {\bibinfo
  {journal} {Europhysics Letters (EPL)}\ }\textbf {\bibinfo {volume} {70}},\
  \bibinfo {pages} {492} (\bibinfo {year} {2005})}\BibitemShut {NoStop}%
\bibitem [{\citenamefont {Manmana},\citenamefont{Hazzard},\citenamefont{Chen},
  \citenamefont{Adrian}\ and\ \citenamefont{Rey}(2011)}]{Manmana_2011}%
  \BibitemOpen
  \bibfield  {author} {\bibinfo {author} {\bibfnamefont {S.~R.}~\bibnamefont
  {Manmana}}, \bibinfo {author} {\bibfnamefont {K.~R.~A.}~\bibnamefont {Hazzard}},
  \bibinfo {author} {\bibfnamefont {G.}~\bibnamefont {Chen}},
  \bibinfo {author} {\bibfnamefont {A.~E.}~\bibnamefont {Feiguin}},
  \ and\ \bibinfo {author} {\bibfnamefont {A.~M.}~\bibnamefont
  {Rey}},\ }\bibinfo {title} {$\rm{SU}\left(N\right)$ magnetism in chains of ultracold alkaline-earth-metal atoms: Mott transitions and quantum correlations}, \href {\doibase 10.1103/PhysRevA.84.043601} {\bibfield
  {journal} {\bibinfo  {journal} {Phys. Rev. A}\ }\textbf {\bibinfo
  {volume} {84}},\ \bibinfo {pages} {043601} (\bibinfo {year}
  {2011})}\BibitemShut {NoStop}%
\bibitem [{\citenamefont {Chen}\ and\ \citenamefont
  {Moukouri}(1996)}]{chen_numerical_1996}%
  \BibitemOpen
  \bibfield  {author} {\bibinfo {author} {\bibfnamefont {L.}~\bibnamefont
  {Chen}}\ and\ \bibinfo {author} {\bibfnamefont {S.}~\bibnamefont
  {Moukouri}},\ }\bibinfo {title} {Numerical renormalization-group study of the one-dimensional $t$-$J$ model}, \href {\doibase 10.1103/PhysRevB.53.1866} {\bibfield
  {journal} {\bibinfo  {journal} {Phys. Rev. B}\ }\textbf {\bibinfo
  {volume} {53}},\ \bibinfo {pages} {1866} (\bibinfo {year}
  {1996})}\BibitemShut {NoStop}%
\bibitem [{\citenamefont {Mahan}(2000)}]{mahan_many-particle_2000}%
  \BibitemOpen
  \bibfield  {author} {\bibinfo {author} {\bibfnamefont {G.~D.}\ \bibnamefont
  {Mahan}},\ }\href@noop {} {\emph {\bibinfo {title} {Many-particle
  physics}}},\ \bibinfo {edition} {3rd}\ ed.,\
  (\bibinfo  {publisher} {Kluwer Academic/Plenum Publishers},\ \bibinfo {year} {2000})\BibitemShut {NoStop}%
\bibitem [{\citenamefont {Wigner}(1934)}]{wigner_interaction_1934}%
  \BibitemOpen
  \bibfield  {author} {\bibinfo {author} {\bibfnamefont {E.}~\bibnamefont
  {Wigner}},\ }\bibinfo {title} {On the Interaction of Electrons in Metals}, \href {\doibase 10.1103/PhysRev.46.1002} {\bibfield  {journal}
  {\bibinfo  {journal} {Phys. Rev.}\ }\textbf {\bibinfo {volume} {46}},\
  \bibinfo {pages} {1002} (\bibinfo {year} {1934})}\BibitemShut {NoStop}%
\bibitem [{\citenamefont {Seidel}\ and\ \citenamefont
  {Lee}(2004)}]{seidel_flux_2004}%
  \BibitemOpen
  \bibfield  {author} {\bibinfo {author} {\bibfnamefont {A.}~\bibnamefont
  {Seidel}}\ and\ \bibinfo {author} {\bibfnamefont {D.-H.}\ \bibnamefont
  {Lee}},\ }\bibinfo {title} {Flux Period, Spin Gap, and Pairing in the One-Dimensional $t$-$J$-$J'$ Model}, \href {\doibase 10.1103/PhysRevLett.93.046401} {\bibfield
  {journal} {\bibinfo  {journal} {Phys. Rev. Lett.}\ }\textbf {\bibinfo
  {volume} {93}},\ \bibinfo {pages} {046401} (\bibinfo {year}
  {2004})}\BibitemShut {NoStop}%
\bibitem [{\citenamefont {Seidel}\ and\ \citenamefont
  {Lee}(2005)}]{seidel_luther-emery_2005}%
  \BibitemOpen
  \bibfield  {author} {\bibinfo {author} {\bibfnamefont {A.}~\bibnamefont
  {Seidel}}\ and\ \bibinfo {author} {\bibfnamefont {D.-H.}\ \bibnamefont
  {Lee}},\ }\bibinfo {title} {The Luther-Emery liquid: Spin gap and anomalous flux period}, \href {\doibase 10.1103/PhysRevB.71.045113} {\bibfield  {journal}
  {\bibinfo  {journal} {Phys. Rev. B}\ }\textbf {\bibinfo {volume} {71}},\
  \bibinfo {pages} {045113} (\bibinfo {year} {2005})}\BibitemShut {NoStop}%
\bibitem [{\citenamefont {Chetcuti}\ (2022)\citenamefont
  {Chetcuti}, \citenamefont {Haug}, \citenamefont {Kwek},\ and\ \citenamefont
  {Amico}}]{chetcuti_persistent_2022}%
  \BibitemOpen
  \bibfield  {author} {\bibinfo {author} {\bibfnamefont {W.~J.}\ \bibnamefont
  {Chetcuti}}, \bibinfo {author} {\bibfnamefont {T.}~\bibnamefont {Haug}},
  \bibinfo {author} {\bibfnamefont {L.-C.}\ \bibnamefont {Kwek}}, \ and\
  \bibinfo {author} {\bibfnamefont {L.}~\bibnamefont {Amico}},\ }\bibinfo {title} {Persistent current of $\rm{SU}\left(N\right)$ fermions}, \href
  {\doibase 10.21468/SciPostPhys.12.1.033} {\bibfield  {journal}
  {\bibinfo  {journal} {SciPost Phys.}\ }\textbf {\bibinfo {volume} {12}},\
  \bibinfo {pages} {033} (\bibinfo {year} {2022})}\BibitemShut {NoStop}%
\bibitem [{\citenamefont {Chetcuti}\ (2023)\citenamefont
  {Chetcuti}, \citenamefont {Polo}, \citenamefont {Osterloh}, \citenamefont {Castorina}, \ and\ \citenamefont
  {Amico}}]{chetcuti_probe_2023}%
  \BibitemOpen
  \bibfield  {author} {\bibinfo {author} {\bibfnamefont {W.~J.}\ \bibnamefont
  {Chetcuti}}, \bibinfo {author} {\bibfnamefont {J.}~\bibnamefont {Polo}},
  \bibinfo {author} {\bibfnamefont {A.}\ \bibnamefont {Osterloh}}, 
  \bibinfo {author} {\bibfnamefont {P.}\ \bibnamefont {Castorina}}, \ and\
  \bibinfo {author} {\bibfnamefont {L.}~\bibnamefont {Amico}},\ }\bibinfo {title} {Probe for bound states of $\rm{SU}\left(3\right)$ fermions and colour deconfinement}, \href
  {\doibase 10.1038/s42005-023-01256-3} {\bibfield  {journal}
  {\bibinfo  {journal} {Commun. Phys.}\ }\textbf {\bibinfo {volume} {6}},\
  \bibinfo {pages} {128} (\bibinfo {year} {2023})}\BibitemShut {NoStop}%
\bibitem [{\citenamefont {Lecheminant}\ (2005)\citenamefont
  {Lecheminant}, \citenamefont {Boulat}, \ and\ \citenamefont
  {Azaria}}]{lecheminant_confinement_2005}%
  \BibitemOpen
  \bibfield  {author} {\bibinfo {author} {\bibfnamefont {P.}\ \bibnamefont
  {Lecheminant}}, \bibinfo {author} {\bibfnamefont {E.}~\bibnamefont {Boulat}},
  \ and\ \bibinfo {author} {\bibfnamefont {P.}~\bibnamefont {Azaria}},\ }\bibinfo {title} {Confinement and Superfluidity in One-Dimensional Degenerate Fermionic Cold Atoms}, \href
  {\doibase 10.1103/PhysRevLett.95.240402} {\bibfield  {journal}
  {\bibinfo  {journal} {Phys. Rev. Lett.}\ }\textbf {\bibinfo {volume} {95}},\
  \bibinfo {pages} {240402} (\bibinfo {year} {2005})}\BibitemShut {NoStop}%
\bibitem [{\citenamefont {Lecheminant}\ (2008)\citenamefont
  {Lecheminant}, \citenamefont {Azaria}, \ and\ \citenamefont
  {Boulat}}]{lecheminant_competing_2008}%
  \BibitemOpen
  \bibfield  {author} {\bibinfo {author} {\bibfnamefont {P.}\ \bibnamefont
  {Lecheminant}}, \bibinfo {author} {\bibfnamefont {P.}~\bibnamefont {Azaria}},
  \ and\ \bibinfo {author} {\bibfnamefont {E.}~\bibnamefont {Boulat}},\ }\bibinfo {title} {Competing orders in one-dimensional half-integer fermionic cold atoms: A conformal field theory approach}, \href
  {\doibase 10.1016/j.nuclphysb.2007.12.034} {\bibfield  {journal}
  {\bibinfo  {journal} {Nuclear Physics B}\ }\textbf {\bibinfo {volume} {798}},\
  \bibinfo {pages} {443} (\bibinfo {year} {2008})}\BibitemShut {NoStop}%
\bibitem [{\citenamefont {Capponi}\ (2008)\citenamefont
  {Capponi}, \citenamefont {Roux}, \citenamefont {Lecheminant}, \citenamefont {Azaria}, \citenamefont {Boulat},
   \ and\ \citenamefont {White}}]{capponi_molecular_2008}%
  \BibitemOpen
  \bibfield  {author} {\bibinfo {author} {\bibfnamefont {S.}\ \bibnamefont
  {Capponi}}, \bibinfo {author} {\bibfnamefont {G.}\ \bibnamefont
  {Roux}}, \bibinfo {author} {\bibfnamefont {P.}\ \bibnamefont
  {Lecheminant}}, \bibinfo {author} {\bibfnamefont {P.}~\bibnamefont {Azaria}}, 
  \bibinfo {author} {\bibfnamefont {E.}~\bibnamefont {Boulat}},
  \ and\ \bibinfo {author} {\bibfnamefont {R.}~\bibnamefont {White}},\ }\bibinfo {title} {Molecular superfluid phase in systems of one-dimensional multicomponent fermionic cold atoms}, \href
  {\doibase 10.1103/PhysRevA.77.013624} {\bibfield  {journal}
  {\bibinfo  {journal} {Phys. Rev. A}\ }\textbf {\bibinfo {volume} {77}},\
  \bibinfo {pages} {013624} (\bibinfo {year} {2008})}\BibitemShut {NoStop}%
\bibitem [{\citenamefont {Arovas}(2008)}]{arovas_simplex_2008}%
  \BibitemOpen
  \bibfield  {author} {\bibinfo {author} {\bibfnamefont {D.~P.}\ \bibnamefont
  {Arovas}},\ }\bibinfo {title} {Simplex solid states of $\rm{SU}\left(N\right)$ quantum antiferromagnets}, \href {\doibase 10.1103/PhysRevB.77.104404} {\bibfield
  {journal} {\bibinfo  {journal} {Phys. Rev. B}\ }\textbf {\bibinfo {volume} {77}},\ \bibinfo {pages} {104404}
  (\bibinfo {year} {2008})}\BibitemShut {NoStop}%
\end{thebibliography}
%
\end{document}